\documentclass[
   pre,aps,
  amsfonts,
  amssymb,
  amsmath,
   superscriptaddress,
   eqsecnum,
   showpacs,
   preprintnumbers,
   byrevtex]{revtex4-1}[10pt]
\usepackage{color}
\usepackage{morefloats}
\usepackage{graphicx}
\usepackage{subfigure}
\usepackage{physics}
\begin{document}
\newcommand{\vphi}{\varphi}
\newcommand{\bq}{\begin{equation}}
\newcommand{\be}{\begin{equation}}
\newcommand{\ba}{\begin{eqnarray}}
\newcommand{\bea}{\begin{eqnarray}}
\newcommand{\eq}{\end{equation}}
\newcommand{\ee}{\end{equation}}
\newcommand{\ea}{\end{eqnarray}}
\newcommand{\eea}{\end{eqnarray}}
\newcommand{\tchi} {{\tilde \chi}}
\newcommand{\tA} {{\tilde A}}
\newcommand{\pstar}{\mbox{$\psi^{\ast}$}}
\newcommand {\bPsi}{{\bar \Psi}}
\newcommand {\bpsi}{{\bar \psi}}
\newcommand{\sn} {\text{sn}}
\newcommand{\cn} {\text{cn}}
\newcommand{\dn} {\text{dn}}
\newcommand{\s} {\textbf{s}}
\newcommand{\tw}{\tilde w} 
\newcommand{\rmi} {i}

\newcommand{\Schrodinger}{Schr{\"o}dinger} 
\preprint{nkdv-nmkdv. \today}
\newpage
%\title{Properties of the  and nmKdV equations and the connection of its solutions to those of  1+1 dimensional $\phi^4$ field theory}
\title{ Connection Between the Exact Moving Solutions of the Negative Korteweg-de Vries (nKdV) Equation and the  Negative Modified Korteweg-de Vries (nmKdV) Equation and the Static Solutions of 1+1 Dimensional $\phi^4$ Field Theory   }

\author{Avinash Khare} 
\email{avinashkhare45@gmail.com}
\affiliation{ Physics Department, 
   Savitribai Phule Pune University, 
   Pune 411007, India}
   
\author{Fred Cooper}
\email{cooper@santafe.edu}
\affiliation{Santa Fe Institute, Santa Fe, NM 87501, USA}
\affiliation{Theoretical Division and Center for Nonlinear Studies, 
Los Alamos National Laboratory, Los Alamos, New Mexico 87545, USA}

\author{Avadh Saxena}
\email{avadh@lanl.gov}
\affiliation{Theoretical Division and Center for Nonlinear Studies, 
Los Alamos National Laboratory, Los Alamos, New Mexico 87545, USA}
\vspace{10pt}

\begin{abstract}

The negative order KdV (nKdV)  and  the modified KdV (nmKdV) equations have two
different formulations based on  different hierarchy operators. Both equations 
can be written in terms of a nonlinear differential equation for a field  
$u(x,t)$ which we call the ``Lou form" of the equation. We find that for 
moving solutions of the nKdV equation and the nmKdV equation written in the 
``Lou form" with $u(x,t)  \rightarrow u (x-ct)= u(\xi) $, the equation for 
$u(\xi)$ can be mapped to the equation for the static solutions of the 1+1 
dimensional $\phi^4$ field theory. Using this mapping we obtain a large number 
of solutions of the nKdV  and the nmKdV equation, most of which are new. We also
show that the nKdV equation can be derived from an Action Principle for both of
its formulations. Furthermore, for both forms of the nmKdV
equations as well as for both focusing and defocusing cases, we show that with
a suitable ansatz one can decouple the $x$ and $t$ dependence of the nmKdV field 
$u(x,t)$ and obtain novel solutions in all the cases. We also obtain novel 
rational solutions of both the nKdV  and the nmKdV equations. 
\end{abstract} 
\pacs{} 
\date{\today}
\maketitle

\section{Introduction}
 Soliton bearing nonlinear evolution equations arise in an increasing number of
 physical systems including hydrodynamics, fluid mechanics, nonlinear optics, 
 classical and quantum field theory, etc.  Most of these equations are of 
 positive order such as the usual KdV and the mKdV equations. However, in 
 recent years there has been a concerted effort in understanding the 
 corresponding negative order equations, particularly the integrable ones such 
 as the negative order KdV (denoted as nKdV) (see, for example \cite{Verosky}) 
 and the negative order mKdV (denoted as nmKdV) (see, for example \cite{Lou}). 
 These equations have intriguing properties and soliton solutions that may be 
 amenable to experimental realization.  For instance, the nKdV  equations are 
 gauge-equivalent to the Camassa-Holm \cite{CH, ch2} equation by invoking reciprocal
 transformations. In addition, they are closely related to the Ermakov-Pinney 
 \cite{E-P} system of equations as well as to the Kupershmidt deformation. Many
 of the properties of the nKdV equation are discussed in  the article  of 
 Qiao and Fan \cite{Qiao-Fan}. The  integrable hierarchy also plays an 
 important role in the theory of non-smooth solitons such as peakons (i.e., 
 peaked solitons), e.g. in the Camassa-Holm equation and cuspons (i.e., cusped 
 solitons) found in the Degasperis-Procesi integrable equation \cite{D-P}. 
 Moreover, it is related to the short pulse equation (SPE) \cite{spe}. 

Connections between the nmKdV and SPE, Liouville equation and the sine-Gordon 
equation have been explored in the literature \cite{WTZ}. Akin to the mKdV, breather 
lattice solutions have been obtained for the nmKdV equation \cite{BL}. The negative order
equations can also describe physical phenomena pertinent to dislocation motion 
in crystals and the propagation of ultrashort optical pulses. In two spatial 
dimensions an integrable negative order generalization of the usual 
Kadomtsev-Petviashvili (KP) equation has also been proposed \cite{Wazwaz} and its 
multi-soliton solutions have been obtained. The latter arises in shallow water 
waves as well as in ion-acoustic waves in plasma physics.

As not above, nonlinear equations are playing increasingly important role in 
several areas of science. Among these, integrable nonlinear equations play a 
very special role, see for example \cite{Das}.  After the remarkable paper of Olver 
\cite{Olver}, it became clear that the recursion operator plays an important 
role in the case of the integrable equations in $1 + 1$ dimensions. The 
existence of a recursion operator for any nonlinear evolution equation 
guarantees the integrability of such a nonlinear equation. For the KdV 
equation, Lenard \cite{Lenard} in 1967 discovered the hierarchy 
\bq
v_t= K^n v \,, 
\eq
where
\bq
 K= - (D^3+2(Dv+vD) ) \,, ~~~D= \partial_x.
\eq
%However, the determination of the recursion operator is not an easy task in general. 
Verosky \cite{Verosky}  extended the work of Lenard in the negative direction
using negative powers in the recursion relation for the usual hierarchy. 

% The usual KdV equation can be written in two ways with two different hierarchy operators.
%The first way  is 
%\bq
%v_t = K v
%\eq
%and the hierarchy is 
%\bq
%
%v_t = K^n v
%\eq
In this approach the negative order KdV equation defined using the recursion 
operator $K$ is given by
\bq \label{dv1} 
K v_t = v \,. 
\eq
The hierarchy of equations can be written as  
\bq
v_t=KG_{m-1}= DG_m \,. 
\eq
Setting $G_0=1$ and $G_{-1}=w$ one obtains the nKdV  equation: 
\bq \label{kdv-1}
v_t=w_x, ~~w_{xxx}+4 v w_x +2 v_x w=0.
\eq
Lou \cite{Lou}  introduced the transformation of variables  
 \be\label{2}
w = u^2\,,~~v = -\frac{u_{xx}}{u}\,, 
\ee
which leads to the coupled system
\be \label{eq7} 
v_t = 2 u u_x,~~ u_{xx}+ vu=0.
\ee
This version of the nKdV equation as well as its decoupled version below we 
will denote as nKdV-1 following \cite{Qiao-Fan}. Eqs. (\ref{eq7})
 can be decoupled in two ways. One form is
\bq \label{eq8} 
u u_{xxt} - u_{xx}u_t + 2u^3u_x = 0\,.
\eq
Another way to decouple Eqs. (\ref{eq7}) which we will call the Lou form of the
equation in what follows, is 
\be \label{dec1} 
\left(\frac{u_{xx}} {u} \right)_t + 2 uu_x =0. 
\ee
Note that  Eq. (\ref{eq8}) can be derived from Eq. (\ref{kdv-1}) under the 
transformation
\bq \label{eq9} 
w=u^2, ~~~v= -\frac{u_{xx}.}{u}. 
\eq
One can also write the KdV equation in a second way  \cite{Lou}
\bq \label{eq1} 
v_t = \Phi v_x , ~~ \Phi = K D^{-1} , 
\eq
where the recursion operator $\Phi$ is given by (here we use the notation of Wazwaz  \cite{Wazwaz19})
\bq \label{eq2} 
\Phi(v) = - (\partial_x^2 + 4 v + 2 v_x \partial_x^{-1}).  
\eq
Starting with this form of the KdV equation we now can define the negative order KdV equation (now called nKdV-2) by  the relationship
\bq \label{eq3} 
\Phi v_t = -v_x  \,.
\eq

Turning our attention to the modified KdV equation (mKdV equation), one has that
the mKdV equation appears in two forms.  The focusing case is described by
\be
u_t+6 u^2 u_x + u_{xxx}=0,
\eq
and the defocusing case by
\bq 
u_t-6 u^2 u_x + u_{xxx}=0,
\eq
Both of these can be written in terms of the recursion operator
\bq
u_t = \Phi  u_x \,. 
\eq
 The recursion operator for the focusing (defocusing) form is given in 
 \cite{Choudhuri, Wazwaz-Xu}
  \bq \label{phi} 
  \Phi = (-D^2 \mp  4 u^2 \mp 4 u_x D^{-1}( \cdot u)) \,, 
  \eq
  where the upper sign refers to the focusing case, and $D=\partial_x$. 
 One form of the  mKdV hierarchy is given by 
  \bq
  u_t = \Phi^n u_x.
  \eq
In that approach, the negative order mKdV equation  can be written as 
\bq  
\Phi   u_t = u_x.
\eq
 The recursion operator for the focusing (defocusing)  form is given by 
 Eq. (\ref{phi}). Another form of the nmKdV equation is found from the Lax 
 pair formulation \cite{Wang}  (see below).
 For the attractive (repulsive)  case one obtains a Lou  form: 
\be \label{eqat}
\big (\frac{u_{xt}}{u} \big )_{x} \pm  (2u^2)_t = 0\,, 
\ee
which is equivalent to 
\be \label{eqat2}
u_{xt} \pm 2u \rho_{xt} +\alpha u = 0\,, ~~\rho_{xx} = u^2\,, 
\ee
where $\alpha$ is an arbitrary integration constant. The upper sign is for the 
attractive case. 

For moving solitary waves that depend only on $\xi=x-ct$, the equations for the 
moving waves for both Eq. (\ref{dec1}) and Eq. (\ref{eqat})  can be related to
the static solutions of the $1+1$  dimensional $\phi^4$ 
field theory. This will be one of the main new results of this paper. 
Letting $u=u(x-ct)= u(\xi)$ for  moving solutions, the Lou equation for the 
nKdV equation  becomes
\bq \label{kdvlouxi}
c(\frac{u_{\xi \xi}} {u})_\xi = -(u^2)_\xi.
\eq
Integrating once we get
\bq \label{dvmov1}
u_{\xi \xi}= \frac{1}{c}  (u^3 + \alpha^\prime  u) ,
\eq
where $\alpha^\prime$ is a constant. Since $c$ and $\alpha^{\prime}$ can have 
either sign, one can let $\alpha= \alpha^\prime/c$ and rewrite the equation for
moving solutions as:
\bq \label{dvmov} 
u_{\xi \xi}= \frac{1}{c}  u^3 + \alpha u .
\eq
Instead for the nmKdV equation, Eq. (\ref{eqat}), we obtain for the focusing 
(attractive) nmKdV-1
\bq \label{phi43}
u_{\xi \xi}= -2 u^3 + \alpha u \,,
\eq
and for the defocusing (repulsive) case:
\bq \label{phi44} 
u_{\xi \xi}= 2 u^3 + \alpha u \,, 
\eq
where again $\alpha$ is an integration constant. We see that 
Eqs. (\ref{dvmov}) to (\ref{phi44}) are related to the equations for the 
static solutions of the $1+1$ dimensional $\phi^4$ field theory,
\be\label{16}
\phi_{xx} = a \phi + b \phi^3\,.
\ee

The main thrust of this paper is how to utilize known static solutions of 
the $\phi^4$ field theory to find novel traveling wave solutions for both the 
nKdV-1  and the nmKdV-1 equations. While some of the traveling wave solutions 
of nKdV-1 and nmKdV-1 are known before \cite{Qiao-Fan, qiao}, most of them 
are not known in the literature so far.  A secondary focus of this paper is to show 
that the nKdV equation has a Lagrangian formulation. We also find novel solutions 
of both the focusing and the defocusing nmKdV equation where the $x$ and $t$ 
dependence decouple.

The plan of the paper is as follows. In Sec. II we discuss the properties of 
the nKdV  equation. We give an action formulation for the equation, and discuss
the infinite conservation laws and the symmetries of both forms of the equation. 
 In Sec. III we turn our attention to the properties of the nmKdV equation. We 
 also discuss the infinite number of conservation laws of the nmKdV equation. 
 In Sec. IV we derive the connection between the traveling wave solutions of 
 both the nKdV-1 and the nmKdV-1 (both focusing and defocusing type) and the 
static solutions of the 1+1 dimensional $\phi^4$ field theory (Eq. (\ref{16})).
Using this connection we obtain a large number of traveling
wave solutions of both these equations. We present only a subset of these 
solutions in this section while the rest are presented in the Appendix. 
In Sec. V we discuss the symmetries of the nKdV equation and also obtain its 
rational solutions. In section VI we discuss the symmetries of the nmKdV-1 and
the nmKdV-2 equations for both the focusing and the defocusing cases and obtain
their rational solutions. Besides, we show that for both the nmKdV-1 and 
nmKdV-2 as well as for both focusing and defocusing type, one can decouple
the $x$ and $t$ dependence and obtain special novel solutions of both nmKdV-1 
and nmKdV-2 equations. Finally, in Section VII we point out some of the open 
problems. 

\section{Properties  of the nKdV equation}
In this section we first consider two forms of the nKdV equation and show how 
both can be written in terms of a potential function $\phi(x,t)$. We then show 
that both forms can be put into the Lou form from which it is easy to derive the 
equation for moving solitons. We also show that both forms of the nKdV equation
in terms of the potential function $\phi$ can be derived from an action $S$ 
which leads to conservation of energy and momentum.    

\subsection{Potential form of the nKdV equation} 

Starting from the nKdV equation, Eq. (\ref{kdv-1}),
\bq 
v_t=w_x, ~~w_{xxx}+4 v w_x +2 v_x w=0,
\eq
Wazwaz \cite{Wazwaz} noted that  the decoupling of the $v$ and $w$ equations is
possible by introducing a potential $\phi(x,t)$, satisfying
\be
w = \phi_t, ~~    v= \phi_x  \,. 
\ee
Then one automatically satisfies the equation
\be
v_t=w_x.
\ee
The equation for $w$ then becomes the following equation for $\phi$
\be \label{waz} 
\phi_{xxxt} + 4\phi_x \phi_{tx} + 2 \phi_{xx} \phi_t = 0 \,. 
\ee
This is the potential form of the nKdV-1 equation.
Using  the negative order KdV equation (nKdV-2)  defined as
\bq  
\Phi v_t = -v_x  \,, 
\eq
with
\bq 
\Phi(v) = - (\partial_x^2 + 4 v + 2 v_x \partial_x^{-1}) \,,   
\eq
and letting 
\bq \label{eq4} 
v_t=\tilde{w}_x\,, 
\eq
one obtains
\bq \label{eq5} 
 \tw_{xxx}+4v \tw_x+2v_x \tw=v_x  \,. 
 \eq
% Wazwaz in \cite{Wazwaz19}  let $v(x) = \phi_x$  which allowed him  to write Eq. ( \ref{eq3}) as: 
%  \bq \label{eq6} 
%\phi_{xxxt} + 4 \phi_x \phi_{xt} + 2 \phi_{xx} \phi_t +\phi_{xx} =0
%\eq
%This equation has the property that it  can be derived from a Lagrangian. 
One can eliminate the rhs of Eq. (\ref{eq5}) by letting
\bq \label{tw} 
\tw(x,t) = w(x,t) + \frac{1}{2}
\eq
to obtain
\bq
v_t= w_x;~~  w_{xxx}+4v w_x+2v_x w=0 \,,  
 \eq
which is identical to  Eq. (\ref{kdv-1}). 

Starting from Eq. (\ref{eq5}), Wazwaz \cite{Wazwaz19}  obtained the nKdV-2  
equation in the following potential form 
\bq  \label{phi2eq} 
\phi_{xxxt} + 4 \phi_x \phi_{xt} + 2 \phi_{xx} \phi_t -\phi_{xx} =0
\eq
by introducing a different  potential $\phi$  which is related to $u$ by
\bq \label{eqtwv}
v(x) = \phi_x =-u_{xx}/u;~~   \phi_t=u^2+1/2 .
\eq
This  explicitly shows the connection between the nKdV-1 and the nKdV-2 equation. 
The different recursion operators $K$ and $\Phi$ lead to equations that can be
transformed into one another by Eq. (\ref{tw}). 
%  We have  not found such a possibility when discussing the two nmKdV equation hierarchies below which have similar recursion operators
One interesting aspect of the nKdV-2  form of the equations is that 
Eq. (\ref{phi2eq}) can be derived from an action formalism which allows one to 
identify both the momentum $P$ of the solution as well as its energy $E$. The 
nKdV-1  equation when cast in an action formulation formally has zero 
Hamiltonian. Using  the compatibility equation
\be
\phi_{xt} = \phi_{tx}
\ee
of Eq. (\ref{eqtwv}) for the nKdV-2 equation leads to the same Lou equation for
$u(x,t)$ as for the nkdV-1 equation, Eq. (\ref{dec1}), namely
\bq  
(\frac{u_{xx}}{u})_t + 2u u_x=0.
\eq
So both nKdV-1 and nKdV-2 formulations lead to  the same equation for the 
moving solutions of the form:  $u=u(x-ct)= u(\xi)$. For moving solutions the 
Lou equation leads to Eq. (\ref{dvmov}). 

The KdV equation as well as the nKdV equation have an infinite number of 
conserved quantities $H_n$. Both equations can be derived from
a Lagrangian when written in terms of the field $\phi$ where $v= \phi_x$. 
%The Hamiltonian hierarchy of infinite conserved quantities is given by (see appendix 1 for details) :
\subsection{Action formulation of the  nKdV equations} 
For the nKdV-2 formulation,  using the identification: 
\be
v= \partial_x \phi(x,t), ~~ \tw = \partial_t\phi(x,t) \,, 
\ee
we obtain the equation

\be \label{phieqa}  
\phi_{xxxt}  + 4 \phi_x \phi_{xt} + 2 \phi_{xx} \phi_t - \phi_{xx} =0.
\ee
On the other hand  the potential form of the  nKdV-1  equation is 
\bq
\phi_{xxxt}  + 4 \phi_x \phi_{xt} + 2 \phi_{xx} \phi_t \,. 
\eq
Both  equations for the potential  can be derived from the action:
\bea
&&S =  \int dt dx \left[ \phi_t (\phi_{xxx} + 2 (\phi_x)^2) 
-\gamma \phi_x^2 \right] \nonumber \\
&&\equiv \int dx dt \left[\pi(x,t) \dot \phi(x,t) - \gamma {h} [\phi_x] \right]
=\int L dt \,, 
\eea
where $\gamma=1$ for the nKdV-2 form of the equation and $\gamma=0$ for the 
nKdV-1 form of the equation. When one has terms in the action with only the 
derivative terms  $\phi_x, \phi_t, \phi_{xx} , \phi_{xxx}$, then setting the 
variation of the action with respect to $\phi$ to zero yields the equation 
 \be
- \partial_t \frac{\delta S}{\delta \phi_t} 
- \partial_x \frac{\delta S}{\delta \phi_x} +\partial_{xx} \frac{\delta S}{\delta \phi_{xx}}- \partial_{xxx} \frac{\delta S}{\delta \phi_{xxx}} =0.
\ee
This leads  to 
\be \label{phieqgam}  
\phi_{xxxt}  + 4 \phi_x \phi_{xt} + 2 \phi_{xx} \phi_t - \gamma \phi_{xx} =0 \,.\ee

The conserved Hamiltonian  is
\be 
 H= I_1 = \gamma [ \int  dx { h}   = \int dx \phi_x^2 = \int dx v(x,t)^2],
 \ee
 and the canonical momentum density is
 \be
 \pi(x,t) = v_{xx} + 2 v^2 = \phi_{xxx} + 2 (\phi_x)^2.
 \ee
 That $H$ is conserved can also be seen by writing Eq. (\ref{phieqa}) for the 
 nKdV-2 equation as a continuity equation:

 \be  \label{phia} 
\partial_t (\phi_x^2) +  \partial_x  ( \phi_{xxt} +2 \phi_x \phi_t - \phi_x ) = 0.
\ee
By considering moving waves we can identify $P$ as 
\bq
P=\int dx ( v_x^2- 2v^3 ) = \int dx (\phi_{xx}^2- 2 \phi_x^3) \,, 
\eq
which is also conserved via the equation of motion, Eq. (\ref{phieqa}).

We notice that the action for the nKdV-2 equation ($\alpha=1$) is very similar 
to the action for the KdV equation \cite{Drazin}  with the role of $P$ and $H$ 
being interchanged. To obtain the Lagrangian for traveling wave  solutions we
let
\be
\phi_x =v = f(x-ct) , ~~~ \phi_t= \tw = g(x-ct) \,, 
\ee
which satisfies the equation  
\bq
\phi_{xt}= \phi_{tx} \,, 
\eq
\be
\tw_x= g'(\xi) = - cf'(\xi) = v_t \,, 
\ee
where $\xi =x-ct$. Thus
\be
g(\xi)  = -c f(\xi)  + \alpha, 
\ee
where $\alpha$ is a  constant. Note that $\tw = w+1/2 $, so the same equation 
holds for both the cases with  different integration constants.
We also have
\bq
\pi(\xi) = \phi_{xxx} + 2 (\phi_x)^2= f''(\xi) + 2 f^2(\xi),
\eq
so that
\bq \label{lag}
L =  \int dy \left[c(f'^2 - 2 f^3 ) - f^2 (1- 2 \alpha )  \right].
\eq
This identifies the momentum of the traveling wave as 
\bq
P=  \int dy (f'^2 - 2 f^3)  \,. 
\eq
The conserved energy and momentum can be identified with two of the infinite 
number of conservation laws found by the Miura Transformation \cite{Qiao-Fan, Miura}.  
The equation  for $f$ is determined by Lagrange's equations with $L$ given by 
Eq. (\ref{lag})
\be
  \frac{d}{dy} \frac {\delta L [f]} {\delta f'}  -\frac{\delta L}{\delta f} = 0 \,, 
\ee
so that
\be \label{linear} 
c[ f'' + 3 f^2] +f(1-2\alpha )  = 0 \,. 
\ee
Solutions  are found by assuming:
\be
f= A  \sech^2(\beta \xi) + B \,. 
\ee
For $\alpha=1$  we find two solutions: $B=0$ and $B=1/(3c)$ with $A= 2 \beta^2$.
When $B=0$ then ${1}/{4 \beta^2}$, when   $B=1/(3c)$ we find 
$c=-{1}/{4 \beta^2}$. Thus we get the two solutions
\bq
f= 2 \beta^2  \sech^2(\beta \xi) \,, 
\eq
\be
f= 2 \beta^2  \sech^2(\beta \xi) -\frac{4}{3} \beta^2.
\ee 

Instead when $\alpha=0$ we find two solutions:   $B=0, c=-{1}/{4 \beta^2}$  and 
\bq
f(\xi) = 2 \beta ^2 \text{sech}^2(\beta \xi)\,. 
\eq
For the second solution, $B=-1/(3 c), c={1}/{4 \beta^2}$ and  
\bq
f(\xi) = 2 \beta ^2 \text{sech}^2(\beta \xi)- \frac{4}{3} \beta^2\,. 
\eq
We see the solutions with $\alpha=1$ and $\alpha=0$ are the same. 
Only the solutions with $B=0$ have finite energy and momentum.  
Eq. (\ref{linear}) does {\em not} have a traveling wave solution of the form
$f= 2 m \beta^2 \cn^2(\beta y,m)$ unless $m=1$.  However, there are solutions 
with $B \neq 0$. Assuming
\bq
f(\xi) =
A c~  \text{cn}(\beta \xi |m)^2+B \,, 
\eq
we find for   $\alpha=1$  
\bq
A c = 2 m \beta^2 ,~~c =\frac{1}{4 \beta^2 \sqrt{1-m+m^2}}, ~~B =\frac{2}{3} 
\beta^2 (1-2 m +\sqrt{1-m+m^2}) \,. 
\eq
Note that when we solve the $\phi$ equation or the $f(\xi)$ equation we are 
solving for $v$ and $g$. When we solve the Lou equation on the other hand,
we directly determine the original variable $u(x,t)$. Note that for the nKdV-1 
equation $ \phi_t= u^2$ and for the nKdV-2,  $\phi_t =u^2+1/2$. 

\subsection{Conservation laws of the  equation}
As shown in Qiao and Fan  \cite{Qiao-Fan},
the  equation possesses infinitely many conservation laws
\bq
F_{n,t} +G_{n,x}  =0, ~~n=1,2, \ldots, 
\eq
where the conserved densities $F_n$ are recursively given by recursion formulas
\ba
F_0 & =&v_{xx} - v^2, ~~F_1 = -v_{xxx}  +2v v_{xx} , \nonumber \\
F_n  & =&I_{n~ xx} - \sum\limits_{k=0}^{n} I_k I_{n-k} \sum\limits_{k=0}^{n-2} I_k I_{n-2-k~x} , ~~n=2,3, \ldots.
\ea

The  fluxes $G_n$ are given by
\ba
G_0 & =&2 w I_0 =2w v, ~~G_1 =2w I_1=   -2w v_x \,,  \nonumber \\
G_n &= &2 w I_n,  n=2,3, \ldots.
\ea
%Note that $w, v$ satisfy the coupled  Eq. (\ref{eq7}).

\section{Properties of the nmKdV equations}
  
We first show that for both the attractive and the repulsive nmKdV-1 equations 
one can obtains a Lou type of equation which then allows one to easily find the 
traveling wave solutions. We also discuss the fact that the nmKdV equation has 
an infinite number of conservation laws. 

The mKdV equation has a bi-Hamiltonian structure as well as a Lagrangian formulation \cite{Choudhuri}.
Writing 
\bq
u= -w_x,
\eq	
then 
\bq
u_{t} =-w_{xt} = -w_{xx}= -P[w_x] .
\eq
If one assumes  $w_x$ vanishes at $\pm \infty$, then
\bq
w_x= w_t.
\eq
Now from
\bq 
u_t = \Lambda_m u_x,
\eq
we have
\ba  \label{weq1}
-w_{xt} &&=  (-D^2 \mp  4w_{xx}  w_x^2 \mp  4 w_{x x} D^{-1}( \cdot w_x))w_{xx}  \nonumber \\
&&=-w_{xxxx} \mp 6 w_{xx} w_x^2 .
\ea
Integrating over $x$ one obtains
\bq
-w_t = -w_{xxx} \mp 2 w_x^3.
\eq
Consider the action: 
\bq
S= \int dx dt [ \alpha w_x w_t + \beta w_x w_{xxx} + \gamma w_x^4] \,. 
\eq
Minimizing the action, we obtain Eq. (\ref{weq1}) by choosing
\bq
\alpha=1/2, ~~\beta=-1/2, ~~\gamma=\mp 1/2 \,, 
\eq
so that the Hamiltonian is given  by 
\bq
H =  1/2 \int dx[ u u_{xx}  \pm u^4]  \,. 
\eq

The negative order mKdV equation can be written as 
\bq  
\Phi   u_t = u_x.
\eq
 The recursion operator for the focusing (defocusing) form is given in 
 \cite{Wazwaz-Xu}
  \bq
  \Phi= (-D^2 \mp  4 u^2 \mp 4 u_x D^{-1}( \cdot u)),
  \eq
  where the upper sign refers to the focusing case.

The nmKdV equation was studied by several researchers \cite{Lou,Wang,qs,gfm,amf,waz1,waz2}. There are two variants of the nmKdV equation that have been proposed in 
the literature and both are integrable. The first one (hereafter we call it as
nmKdV-1) is given by
\be\label{10}
u u_{xxt} - u_x u_{xt} +4 u^3 u_t = 0\,,
\ee
in the focusing case while 
\be\label{11}
u u_{xxt} - u_x u_{xt} -4 u^3 u_t = 0\,,
\ee
in the defocusing case. On the other hand, using the recursion operator for
the mKdV case, Wazwaz \cite{waz1} obtained the following nmKdV equation
(hereafter we call it as nmKdV-2)
\be\label{12}
u_x u_{xxxt} + 4 u^2 u_x u_{xt} +12 u u_{x}^{2} u_t -u_{xx} u_{xxt} 
-4 u^2 u_{xx} u_t = 0\,,
\ee
in the focusing case while 
\be\label{13}
u_x u_{xxxt} - 4 u^2 u_x u_{xt} -12 u u_{x}^{2} u_t -u_{xx} u_{xxt} 
+4 u^2 u_{xx} u_t = 0\,,
\ee
in the defocusing case. Subsequently,  a few soliton, traveling wave and 
breather solutions have been obtained for these nmKdV equations 
\cite{Wang,waz1}. 

The nmKdV-2 is obtained from the hierarchy  operator $\Phi$.
The negative order mKdV-2  equations are given by
\bq
\Phi u_t = u_x \,. 
\eq
Here $\Phi$ is given by  \cite{Wazwaz-Xu}
\bq 
\Phi =- D^2  \mp 4u^2  \mp 4 u_x D^{-1}  ( \cdot u) ,
\eq
where the upper sign is for the focusing branch. 
Here
\bq
D^{-1}  ( \cdot u)f(x) = D^{-1}  ( f(x) \cdot u) \,, 
\eq
so we obtain 
%  \bq
% [ -D^2  \mp 4u^2  \mp 4 u_x D^{-1}  ( \cdot u)]  u_t = u_x
%  \eq
%  which yields
  \bq \label{ugga5} 
  -u_{xxt}  \mp u^2 u_t \mp 4 u_x \partial^{-1} (u u_t) = u_x.
  \eq
  
  Note that unlike the mKdV equation where we had the term
 \bq
   \partial^{-1} (u u_x)  = u^2/2 \,, 
   \eq
   we now have to set  
 $ (u^2/2)_t  = w_x$
 to undo the inverse derivative with respect to $x$. 
Another strategy is to  differentiate Eq. (\ref{ugga5}) with respect to $x$
and use
\bq \label{ugga6} 
 \partial^{-1}  ( u \cdot  u_t) =  \frac{u_x +u_{xxt} \mp  4 u^2 u_t}{4 u_x}  \,. 
 \eq
 
Doing this we obtain
 \bq \label{ju1} 
 -u_x u_{xxxt} \mp 4 u^2 u_x u_{xt} \mp 12 u u_x^2 u_t + u_{xx} u_{xxt}  \pm 4 u^2 u_{xx}u_t =0.
 \eq
 We can get a slightly simpler equation by using $w$ 
  \bq \label{mugga5} 
  -u_{xxt}  \mp u^2 u_t \mp 4 u_x w= u_x.
  \eq
Then choose   $\tw$ so that 
 \bq \label{mugga6} 
 \mp 4 u_x w- u_x=  \mp 4 u_x \tw \,. 
  \eq
Then we have 
  \bq \label{mugga7} 
  -u_{xxt}  \mp u^2 u_t \mp 4 u_x  \tw= 0.
  \eq
We can now differentiate Eq. (\ref{mugga7}) with respect to $x$  and then use 
Eq. (\ref{mugga5}) to eliminate $\tw$.  Using $w_x=\tw_x= u u_t$, this leads to:
   \bq
 \pm u_{xxxt}  + 2 u u_x u_t+ u^2 u_{xt} +4 u_{xx} \tw + 4 u_x u u_t =0 \,, 
 \eq
where 
\bq 
4 u_{xx} \tw = \mp u_{xxt} u_{xx}/u_x - u^2 u_t u_{xx}/u_x \,, 
\eq
so we obtain 
\bq \label{aarg} 
\pm u_{xxxt}u_{x} + \mp u_{xxt} u_{xx} + u^2 u_t u_{xx} + 4 u_x u u_t =0 \,. 
  \eq 
Unlike the mKdV case or the nKdV case,  we have not been able to find a Lagrangian that leads to these equations. 
  
Using the Lax pair approach \cite{Wang} for obtaining the infinite set of 
conservation laws, one gets a simpler set of equations which we
denote as the nmKdV-1 equations. 
%(see Eq.(\ref{nmkdvsimp}) below) . 
For the focusing (defocusing)  case one obtains: 
\be\label{4at}
\big (\frac{u_{xt}}{u} \big )_{x} \pm  (2u^2)_t = 0\,, 
\ee
which is equivalent to 
\be\label{4.2}
\rho_{xx} = u^2\,,~~u_{xt} \pm 2u \rho_{xt} +\alpha u = 0\,,
\ee
where $\alpha$ is an arbitrary integration constant. The upper sign is for the 
attractive case. 

%The corresponding equation for the repulsive  nmKdV equation is
%\be\label{4rep}
%\big (\frac{u_{xt}}{u} \big )_{x} - (2u^2)_t = 0\,,
%\ee
%which is equivalent to 
%\be\label{4.6}
%\rho_{xx} = u^2\,,~~u_{xt} -2u \rho_{xt} +\alpha u = 0\,.
%\ee
In analogy with Eq. (\ref{ju1}) we can write these equations as: 
 \be  \label{4.3at}
u u_{xxt} - u_x u_{xt} \pm  4 u^3 u_t = 0\,.
\ee
The upper sign is for the attractive (focusing) case.

We have not been able to find the simple transformation from one form i.e. Eq. (\ref{4.3at}) to the other i.e. Eq. (\ref{ju1}), nor have we been able to find a Lagrangian for either form of the nmKdV equations.

The  equation in Lou form, Eq. (\ref{dec1}), is quite similar to Eq. (\ref{4at})
%\bq
%(\frac {u_{xx}}{u})_t  +2 u u_x = 0.
%\eq
which leads to the important fact that for the moving solitary waves that 
depend only on $\xi=x-ct$, the equations for the moving waves will be related.  

\subsection{Conservation Laws  for nmKdV-1}

The nmKdV equations have an infinite number of conservation laws.
These conservation laws can be determined from the Lax pair. The formalism is 
discussed in detail in \cite{Wang}.
Here we sketch the formalism for the attractive case, the repulsive case can be
obtained from the substitution $u \rightarrow \rmi u$.
From the Lax pair
\bq
\Phi_x =  A \Phi, ~~~ \Phi_t = B \Phi,
\eq
where  
\bq
\Phi =(\Phi_1, \Phi_2) _T,
\eq
and
\bq
A=
 \left(  \begin{array}{cc}
\lambda & iu \\
iu & - \lambda  \end{array}  \right) \,, 
\eq
\bq
B=
 \frac{1}{2 \lambda} \left(  \begin{array}{cc}
-\frac{\alpha}{2} - \rho_{xt} & -i u_t \\
i u_t  & \frac{\alpha}{2} +\rho_{xt}\end{array}  \right) \,. 
\eq
Then setting $ \Gamma = {\Phi_2}/{\Phi_1}$ one finds that $\Gamma$ obeys a continuity equation 
\bq \label{cont2} 
\frac{d}{dt} (iu \Gamma) = \frac{1}{2 \lambda} \frac{d}{dx} ( - \rho_{xt} - i u_t \Gamma),
\eq
as well as the equation:
\bq
\Gamma_x= iu - 2 \lambda \Gamma - iu \Gamma^2.
\eq
Assuming  $\Gamma$ has a series in inverse powers of $\lambda$  then leads to a set of recursion relations for the coefficients $\Gamma_n$. One finds:
\bea
\Gamma_1 && = \frac{i}{2} u,~~~  \Gamma_2 = -\frac{i}{4} u_x \,, \nonumber \\
\Gamma_n &&= - \frac{1}{2}\Gamma_{n-1,x}- \frac{i}{2} u \sum_{j+k=n-1} \Gamma_j \Gamma_k= 0, ~~n \ge 3.
\ea

Using the continuity equation, Eq. (\ref{cont2}), one then obtains Eq. (\ref{4at}), namely
\bq \label{nmkdvsimp}
(2 u^2)_t + (\frac {u_{xt}}{u})_x=0 \,. 
\eq
%We One can write this as
%\bq
%\rho_{xx}=u^2,~~ u_{xt} + 2 u \rho_{x,t} + \alpha u  =0
%\eq
%Note that $\alpha$ is an arbitrary integration constant. 
One also obtains the recursion relation 
\bq
(2u \Gamma_n)_t + (u_t \Gamma_{n-1})_x = 0, ~~n \ge2.
\eq
This leads to an infinite set of conservation laws. The first three conservation laws are (the upper sign is for the attractive case): 
\bq
M_1 = 8 \int u^2 dx;~~M_2= 2 \int (u^4 \mp u_x^2) dx;~~M_3= \int  (u^6  \mp 5 u^2 u_x^2 + \frac{1}{2} u_{xx}^2) dx.
\eq

\section{\bf  Connection between traveling wave solutions of nKdV-1 and nmKdV-1
equations and static solutions of $\phi^4$ field theory}

The Lagrangian of 1+1 dimensional $\phi^4$ field theory is 
\bq
L= \int dx (\partial_\mu \phi  \partial^\mu \phi  \mp m^2 \phi^2/2 -\lambda \phi^4/4) \,, 
\eq
whose time independent solutions obey
\bq \label{phieq}
\phi_{xx} = \pm m^2  \phi + \lambda \phi^3 \equiv  a \phi + b \phi^3.
\eq
We would like to show that the moving solutions  $u(x-ct)=u(\xi)$ of both the
nKdV-1 and the nmKdV-1 equations can be related to the solutions of 
Eq. (\ref{phieq}). However, the conservation laws of the equation are functions 
of $v(x,t)$ and the conservation laws of the nmKdV-1 equation are functions of 
$u(x,t)$. Thus, to determine the finite energy moving solutions of the nKdV-1 
equation one has to first obtain $v(x,t)$ from $u(x,t)$. Thus each static 
solution of the $\phi^4$ field equation leads to different solutions for the 
nKdV-1 and the nmKdV-1 equations if we are interested in solutions with finite 
energy. It was already shown by Qiao and Li \cite{qiao} that for the 
traveling wave solutions of the nKdV-1 equation, the Lou form of the equation 
could be mapped into the solutions of the $\phi^4$ field theory. Here we show 
that their observation can be extended to the nmKdV-1 equation and that the 
solutions of Eq. (\ref{phieq}) allow us to determine the moving solutions for 
both the nKdV-1 and the nmKdV-1 equations. 

If we consider the moving solutions of the  nKdV-1 equation with  $\xi = x-ct$,
the Lou form of the  equation is  Eq (\ref{kdvlouxi}) 
%\be\label{3.2}
%c [u u_{\xi \xi \xi} -u_{\xi} u_{\xi \xi}] = 2 u^3 u_{\xi}\,.
%\ee
%On rewriting Eq. (\ref{3.2}) as
\be\label{3.3}
c  \frac{d}{d\xi} (\frac{u_{\xi \xi}}{u}) = 2 u u_{\xi}\,.
\ee
One obtains by integrating over $\xi$
\be\label{3.4}
u_{\xi \xi} = (1/c) u^3 +\alpha u\,,
\ee
which is similar to the static field equation of the symmetric $\phi^4$ 
Eq. (\ref{phieq}). Here $\alpha$ is an integration constant. Note, depending on 
whether the wave is moving to the right or to the left, i.e. if $c > 0$ or 
$c < 0$, the coefficient of the $u^3$ term can be positive or negative. Thus 
we can identify $1/c$ and $\alpha$ with $\phi^4$ coefficients $b$ and $a$,  
respectively. Further, while in the $\phi^4$ case one has a static field
equation, in the nKdV-1 case one is discussing the moving solitary wave 
equation. For the nKdV-1 equation the field that enters the conservation laws 
is
\bq
v= -\frac{u_{xx}} u = -u'' (\xi)/u(\xi)= - u^2/c - \alpha,
\eq
where we have used Eq. (\ref{3.4}).

In terms of $\xi = x-ct$ it is easy to see that the focusing nmKdV-1 
Eq. (\ref{4at}) can be rewritten as
\be\label{3.5}
(\frac{u_{\xi \xi}}{u})_{\xi} + (2 u^2)_{\xi} = 0. 
\ee
Note that Eq. (\ref{3.5}) is independent of $c$. Integrating once one gets
\be\label{3.6}
u_{\xi \xi} = -2 u^3 + \alpha u\,,
\ee
where $\alpha$ is an integration constant. Thus for the focusing nmKdV-1 case
we can identify $-2$ and $\alpha$ with the $\phi^4$ coefficients $b$ and $a$,  
respectively.

On the other hand, the defocusing nmKdV-1 Eq. (\ref{4at}) can be rewritten as
\be\label{3.7}
(\frac{u_{\xi \xi}}{u})_{\xi} - (2 u^2)_{\xi} = 0\,.
\ee
%Note Eq. (\ref{3.7}) is also independent of $c$. 
Integrating once one gets
\be \label{3.8}
u_{\xi \xi} = 2 u^3  + \alpha u\,, 
\ee
where $\alpha$ is an integration constant. Thus, for the defocusing nmKdV-1 case
we can identify $2$ and $\alpha$ with the $\phi^4$ coefficients $b$ and $a$, 
respectively.

Summarizing, in nKdV-1 as well as in both the focusing and the defocusing 
nmKdV-1 cases the integration constant $\alpha$ can be identified with the 
$\phi^4$ coefficient $a$. On the other hand, while the nKdV-1 coefficient $1/c$ 
can be identified with the $\phi^4$ coefficient $b$, for the focusing (defocusing)
nmKdV-1 cases the coefficient $-2$ ($2$) is identified with the $\phi^4$ 
coefficient $b$. Thus all the well known static solutions of the symmetric 
$\phi^4$ field theory are automatically the solutions of the nKdV-1 or either
the attractive or the repulsive nmKdV-1 equations.

We now discuss the various known static solutions of the symmetric $\phi^4$ 
Eq. (\ref{phieq}) and then write down the corresponding solutions of the nKdV-1 
Eq. (\ref{3.4}) and the appropriate nmKdV-1 equation, i.e. either 
Eq. (\ref{3.6}) or (\ref{3.8}). It is worth pointing out that while some of the
traveling wave solutions of the nKdV-1 and the nmKdV-1 are already known 
\cite{qiao}, most of the solutions that we present here are new. In this section we 
present 12 solutions of Eq. (\ref{phieq}) and the corresponding solutions of 
the nKdV-1 and the nmKdV-1 equations. The remaining 20 solutions are given in  
the Appendix.

% s\ubsection{\bf cnoidal traveling wave solutions to Eq. (\ref{phieq}) }

{\bf Solution I} \\
It is well known \cite{aubry, CKMS} that the static $\phi^4$ Eq. (\ref{phieq}) 
admits the pulse-like periodic solution 
\be\label{5.1}
\phi(x) = A \sqrt{m} \cn(\beta x,m)\,,
\ee
provided
\be\label{5.2}
b A^2 = -2 \beta^2\,,~~a = (2m-1)\beta^2\,.
\ee
Thus the corresponding periodic solution of the nKdV-1 Eq. (\ref{3.4}) is
\bq 
u(\xi) = \sqrt{2 |c| } \beta  \sqrt{m} \cn(\beta \xi,m)\, ,
\eq
so that
\be\label{5.3}
v(x) =  -\beta^2  \left[(2m-1)+ 2| c| m \cn^2(\beta \xi,m) \right].
\ee
When $m \rightarrow 1$, we obtain the hyperbolic solution
\be
v(x) = -\beta^2  \left[ 1+  \sech^2(\beta \xi) \right].
\ee

In Fig. (1) we plot $v(\xi)$ of Eq. (\ref{5.3}) vs $\xi$ in case $m = 1/2, 
c = \beta = 1$.
\begin{figure}
\includegraphics[width=0.4\linewidth]{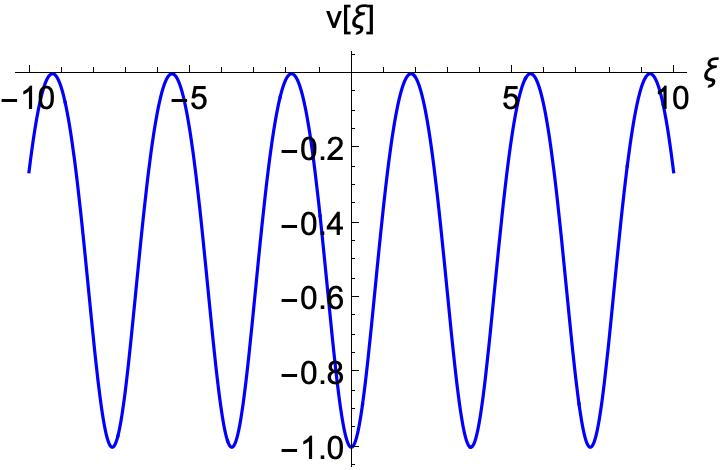}
\caption{The solution $v(\xi)$ of Eq. (\ref{5.3}) vs $\xi$ in case $m = 1/2, c = 
\beta = 1$\,. }
	\label{fig:fig1}
\end{figure}
For the focusing nmKdV-1 Eq. (\ref{3.6}) the $\cn(x,m)$ solution is 
\bq
u(\xi) = \beta  \sqrt{m} \cn(\beta \xi,m) \,. 
\eq
For $m=1$ the solution reduces to the solitary wave solution
\be\label{5.4}
u(\xi) = \beta  \sech (\beta \xi)\,.
\ee
In Fig. (2) we plot $u(\xi)$ of Eq. (\ref{5.4}) vs  $\xi$ in case $m = 1/2, 
\beta = 1$.
\begin{figure}
\includegraphics[width=0.4\linewidth]{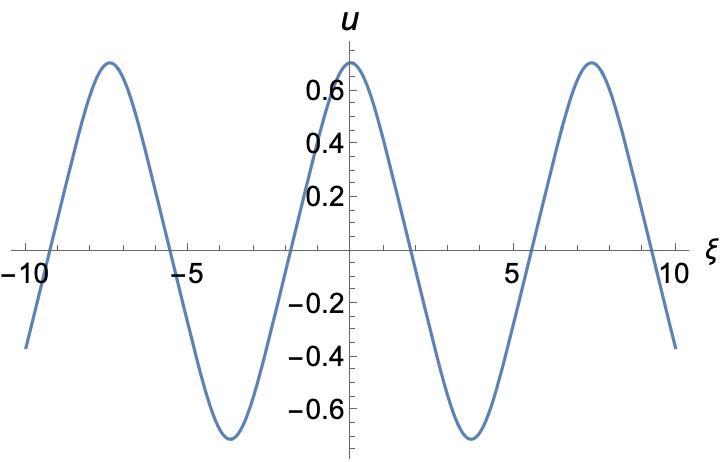}
\caption{The solution $u(\xi)$ of Eq. (\ref{5.4}) vs $\xi$ in case $m = 1/2, 
\beta = 1$\,. }
	\label{fig5}
\end{figure}

{\bf Solution II} \\
The field Eq. (\ref{phieq}) admits  \cite{aubry, CKMS} the periodic solution
\be\label{5.11}
\phi(x) = A\sqrt{m}  \sn(\beta x,m)\,,
\ee
provided
\be\label{5.12}
b A^2 = 2 \beta^2\,,~~a = -(1+m)\beta^2\,. 
\ee
This is a periodic solution except for $m=1$ when it becomes a kink solution. 

It then follows that the nKdV-1 Eq. (\ref{3.4}) has the traveling periodic wave
solution 
\be\label{5.13}
u(\xi) =  \sqrt{2 c} \beta \sqrt{m}  \sn(\beta \xi,m), ~~c > 0\,.
\ee
In Fig. (3) we plot $u(\xi)$ of Eq. (\ref{5.13}) vs  $\xi$ in case $\beta = 1, 
m = 1/2$\,.
\begin{figure}
\includegraphics[width=0.4\linewidth]{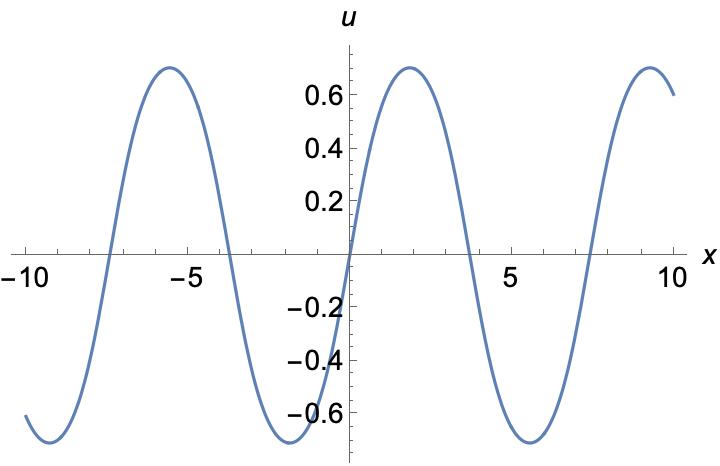}
\caption{The solution $u(x)$ of (\ref{5.13}) vs $\xi$ for $\beta=1, m=1/2$. }
	\label{snx}
\end{figure}
% provided
%\be\label{5.13}
%A^2 = 2c \beta^2\,,~~\alpha = -(1+m)\beta^2\,.
%\ee
Therefore
\bq
u^2(\xi) = 2c \beta^2 m ( \sn(\beta \xi ,m))^2
\eq
and
\be
v(\xi) = -u^2/c - \alpha  = \beta^2 [(m+1)-2 m  ( \sn(\beta \xi ,m))^2] \,. 
\ee
When $m=1$ we obtain the hyperbolic solitary wave solution
\be\label{5.14}
v(x) =2 \beta ^2 \text{sech}^2(\beta  \xi),
\ee
In Fig. (4) we plot $v(\xi)$ of Eq. (\ref{5.14}) vs  $\xi$ in case $\beta = m 
= 1$.  
\begin{figure}
\includegraphics[width=0.4\linewidth]{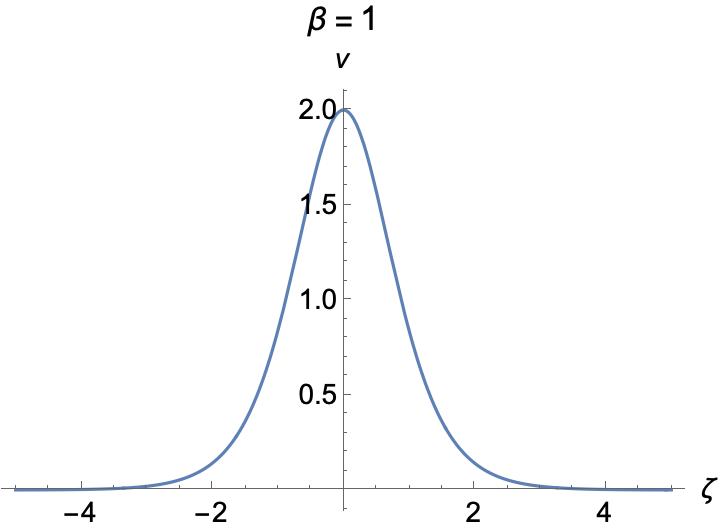}
\caption{The solution $v(\xi)$ of Eq. (\ref{5.14}) vs $\xi$  for $\beta = m=1$. } 
\label{vsech}
\end{figure}

For the defocusing nmKdV-1 Eq. (\ref{3.8}) we instead have the solution
\bq
u(\xi) =  \beta \sqrt{m}  \sn(\beta \xi,m) \,. 
\eq
%provided
%\be\label{5.14}
%A^2 =  \beta^2\,,~~\alpha = -(1+m) \beta^2\,.
%\ee
In the limit $m = 1$, the periodic solution goes over to the usual kink 
solution
\be\label{5.15}
u(\xi) = \beta \tanh(\beta \xi)\,. 
\ee
In Fig. (5) we plot $u(\xi)$ of Eq. (\ref{5.15}) vs  $\xi$ in case $\beta = m 
= 1$.  
\begin{figure}
\includegraphics[width=0.4\linewidth]{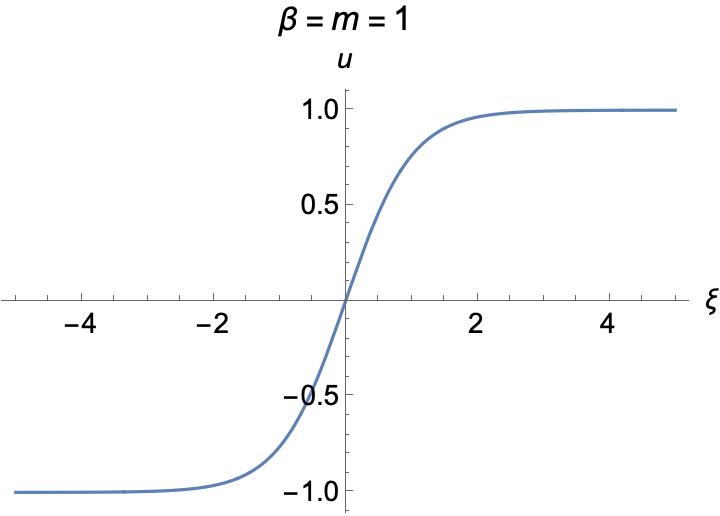}
\caption{The solution $u(\xi)$ of Eq. (\ref{5.15}) vs $\xi$ for $\beta = m=1$. }
\label{utanh}
\end{figure}

{\bf Solution III} \\
Eq. (\ref{phieq}) admits a complex
PT-invariant periodic pulse solutions with $PT$ 
eigenvalue +$1$  \cite{ks1}
\be\label{5.20}
\phi(x) = \sqrt{m} [A\cn(\beta x, m) +i B \sn(\beta x, m)]\,,
\ee
provided 
\be\label{5.21}
B = \pm A\,,~~2bA^2 = -\beta^2\,,~~a = -\frac{(2-m)\beta^2}{2}\,.
\ee

%\subsubsection{\bf Complex  solution}

Hence the nKdV-1 Eq. (\ref{3.4}) has the solution  ($c<0$)
\bq
u(\xi) = \sqrt{\frac{m|c|}{2}}\beta 
[\cn(\beta \xi, m) \pm i \sn(\beta \xi, m)]\,,
\eq
so that 
\bq
v(\xi) = -\frac{\sqrt{m}\beta^2}{2} [\cn(\beta \xi, m) 
\pm i \sn(\beta \xi, m)]^2+ \frac{(2-m)\beta^2}{2} \,. 
\eq

%(\ref{5.20}) provided
%\be\label{5.22}
%2A^2 = |c| \beta^2\,,~~\alpha = -\frac{(2-m)\beta^2}{2}\,.
%\ee
%
%\subsubsection{\bf Complex nmKdV solution}

On the other hand the focusing nmKdV-1 Eq. (\ref{3.6}) has the solution 
\be\label{5.23}
u(\xi) = \frac{\sqrt{m} \beta}{2} [(\cn(\beta \xi, m) \pm i 
\sn(\beta \xi, m)]\,,
\eq
with $\alpha = -\frac{(2-m)\beta^2}{2}$\,. 

In the limit $m = 1$, Eq. (\ref{5.20})  goes  over to the complex
hyperbolic pulse solution with PT-eigenvalue +$1$
\be\label{5.24b}
\phi(x) = A \sech(\beta x) +i B \tanh(\beta x)\,,
\ee
provided relations (\ref{5.21}) with $m = 1$ are satisfied.

{\bf Solution IV}\\

Eq. (\ref{phieq})  also admits a complex
PT-invariant kink solution with $PT$ eigenvalue -$1$ \cite{ks1}
\be\label{4.22}
\phi(x) = \sqrt{m}[A \sn(\beta x, m) +i B \cn(\beta x, m)]\,,
\ee
provided 
\be\label{4.23}
B = \pm A\,,~~2bA^2 =  \beta^2\,,~~a = -\frac{(2-m)\beta^2}{2}\,.
\ee
In the hyperbolic limit $m = 1$, the solution (\ref{4.22}) goes over to the
complex PT-invariant hyperbolic solution with PT-eigenvalue $-1$
\be\label{4.24}
\phi(x) = A  \sech(\beta x) +i B \tanh(\beta x)\,.
\ee

Hence the corresponding complex PT-invariant kink solution of nKdV-1 Eq. 
(\ref{3.4}) is ($c > 0$ )
\be\label{4.25}
u(\xi) = \sqrt{\frac{mc}{2}} \beta[(\cn(\beta \xi, m) \pm   
i\sn(\beta \xi, m)]\,.
\ee

On the other hand, the corresponding complex PT-invariant kink solution of the
defocusing nmKdV-1 Eq. (\ref{3.8}) is
\be\label{4.26}
u(\xi) = \frac{\sqrt{m} \beta}{2} [\cn(\beta \xi, m) \pm i\sn(\beta \xi, m)]\,.
\ee

{\bf Solution V}\\

Remarkably, we find that Eq. (\ref{phieq}) also admits
a real superposed pulse solution \cite{ks2}
\be\label{4.27}
\phi(x) = A \dn(\beta x, m) + B \sqrt{m} \cn(\beta x, m)\,,
\ee
provided
\be\label{4.28}
B = \pm A\,,~~ 2bA^2  = -\beta^2\,,~~a = -\frac{(1+m)\beta^2}{2}\,.
\ee

Hence the corresponding superposed pulse solution of the nKdV-1 Eq. (\ref{3.4})
is ($c < 0$)
\be\label{4.29}
u(\xi) = \sqrt{\frac{|c|}{2}} \beta [\dn(\beta \xi,m) 
\pm \sqrt{m}  \cn(\beta \xi,m)]\,.
\ee

On the other hand, the corresponding superposed pulse solution of the 
focusing nmKdV-1 Eq. (\ref{3.6}) is
\be\label{4.30}
u(\xi) = \frac{\beta}{2} [\dn(\beta \xi,m) 
\pm \sqrt{m}  \cn(\beta \xi,m)]\,.
\ee

{\bf Solution VI}\\

One finds that \cite{ks3} 
\be\label{4.31}
\phi(x) = \frac{[A\sin(\beta x)+iB]}{D+\cos(\beta x)}\,,
~~D > 1\,,
\ee
is an exact complex PT-invariant periodic solution with PT-eigenvalue 
$-1$ of $\phi^4$ Eq. (\ref{phieq}) provided
\be\label{4.32}
2 b A^2 = \beta^2\,,~~2 b B^2 = (D^2-1)\beta^2\,,~~
a = \frac{\beta^2}{2}\,.
\ee

Hence the corresponding superposed pulse solution of the nKdV-1 Eq. (\ref{3.6})
is ($c > 0$)
\be\label{4.33}
u(\xi) = \frac{\sqrt{\frac{c}{2}} \beta [\sin(\beta \xi) 
+ i\sqrt{D^2-1}]}{D+\cos(\beta \xi)}\,.
\ee

On the other hand, the corresponding superposed pulse solution of the 
defocusing nmKdV-1 Eq. (\ref{3.8}) is
\be\label{4.34}
u(\xi) = \frac{\frac{\beta}{2} [\sin(\beta \xi) 
+i\sqrt{D^2-1}]}{D +\cos(\beta \xi)}\,.
\ee

{\bf Solution VII} \\

Eq. (\ref{phieq}) also admits the periodic superposed solution 
\cite{ks1}
\be\label{4.35}
\phi(x) = A \dn(\beta x,m) +\frac{B\sqrt{1-m}}{\dn(\beta x, m)}\,,~~
0 < m < 1\,,
\ee 
provided
\be\label{4.36}
B = \pm A\,,~~ bA^2  = -2\beta^2\,,~~a = (2-m \pm 6\sqrt{1-m})\beta^2\,.
\ee
Note that the two $\pm$ signs in Eq. (\ref{4.36}) are correlated.

Hence the corresponding superposed pulse solution of the nKdV-1 Eq. (\ref{3.4})
is ($c < 0$)
\be\label{4.37}
u(\xi) = \sqrt{2|c|}\beta[\dn(\beta \xi,m) \pm 
\frac{\sqrt{1-m}}{\dn(\beta \xi, m)}]\,.
\ee

On the other hand, the corresponding superposed pulse solution of the 
focusing nmKdV-1 Eq. (\ref{3.6}) is
\be\label{4.38}
u(\xi) = \beta[\dn(\beta \xi,m) \pm 
\frac{\sqrt{1-m}}{\dn(\beta \xi, m)}]\,.
\ee

{\bf Solution VIII} \\

It is easy to check that \cite{ks3}
\be\label{4.39}
\phi(x) = \frac{A\sqrt{m}\sn(\beta x,m)}{D+\dn(\beta x,m)}\,,~~D > 0\,,
\ee
is an exact periodic pulse solution of the symmetric $\phi^4$ Eq. (\ref{phieq}) 
provided
\be\label{4.40}
D = 1\,,~~2 b A^2 =  \beta^2\,,~~a = -(2-m)\frac{\beta^2}{2}\,.
\ee

Hence the corresponding superposed pulse solution of the nKdV-1 Eq. (\ref{3.4})
is ($c > 0$)
\be\label{4.41}
u(\xi) = \frac{\sqrt{\frac{c}{2}} \beta \sn(\beta \xi,m)}
{D+\dn(\beta \xi,m)}\,.
\ee

On the other hand, the corresponding superposed pulse solution of the 
defocusing nmKdV-1 Eq. (\ref{3.8}) is
\be\label{4.42}
u(\xi) = \frac{\frac{\beta}{2}\sn(\beta \xi,m)}{D+\dn(\beta \xi,m)}\,.
\ee

{\bf Solution IX} \\

The $\phi^4$ field Eq. (\ref{phieq})  admits 
the periodic solution \cite{ks4}
\be\label{4.43}
\phi(x) = \frac{A \dn(\beta x,m) \cn(\beta x,m)}{1+B\cn^2(\beta x,m)}\,,
~~B > 0\,,
\ee
provided
\bea\label{4.44}
&&0 < m < 1\,,~~B = \frac{\sqrt{m}}{1-\sqrt{m}}\,,
~~a = -[1+m+6\sqrt{m}]\beta^2 < 0\,, \nonumber \\
&&b A^2 = \frac{8 \sqrt{m} \beta^2}{(1-\sqrt{m})^2}\,.
\eea
On using the identity \cite{as}
\be\label{4.45}
\sn(y+\Delta,m)-\sn(y-\Delta,m) = \frac{2\cn(y,m) \dn(y,m) \frac{\sn(\Delta,m)}
{\dn^2(\Delta,m)}}{1+B\cn^2(y,m)}\,,~~B = \frac{m \sn^2(\Delta,m)}
{\dn^2(\Delta,m)}\,,
\ee
one can re-express the periodic solution IX as a periodic kink-antikink-  
lump, i.e.
\be\label{4.46}
\phi(x) = \frac{\sqrt{2m} \beta}{\sqrt{b}} 
\bigg [\sn(\beta x +\Delta, m) -\sn(\beta x- \Delta, m) \bigg ]\,. 
\ee
Here $\Delta$ is defined by $\sn(\sqrt{m}\Delta,1/m) = \pm m^{1/4}$,
where use has been made of the identity
\be\label{4.47}
\sqrt{m} \sn(y,m) = \sn(\sqrt{m} y,1/m)\,. 
\ee

It then follows that the nKdV-1 Eq. (\ref{3.4}) admits the solution 
\be\label{4.48}
u(\xi) = \frac{A \dn(\beta \xi,m) \cn(\beta \xi,m)}{1+B\cn^2(\beta \xi,m)}\,,
~~B > 0\,,~~0 < m < 1\,,
\ee
provided
\be\label{4.49}
B = \frac{\sqrt{m}}{1-\sqrt{m}}\,,~~A^2 = c \frac{8 \sqrt{m} \beta^2}
{(1-\sqrt{m})^2}\,.
\ee
In Fig. (6) we plot $u(\xi)$ of Eq. (\ref{4.48}) vs  $\xi$ in case $\beta = c = 1$, 
$m = 1/2$.  
\begin{figure}
\includegraphics[width=0.4\linewidth]{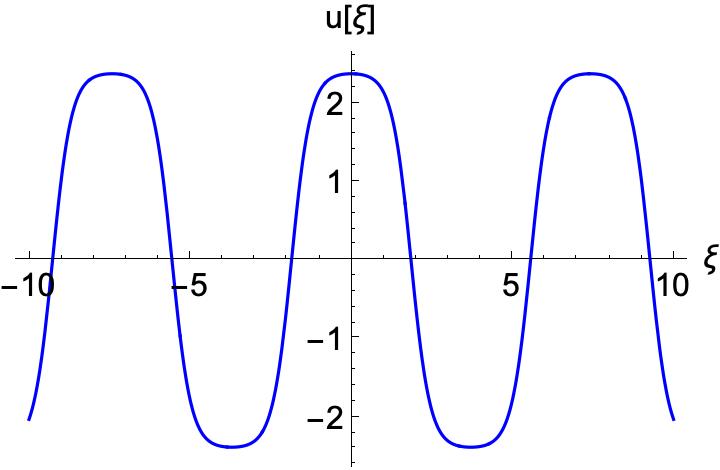}
\caption{The solution of $u(\xi)$ of Eq. (\ref{4.48})  vs $\xi$  for $\beta = c = , 
1m=1/2$. }
\label{sol9}
\end{figure}

It follows that the defocusing nmKdV-1 Eq. (\ref{3.8}) also admits the solution 
(\ref{4.48}) provided
\be\label{4.50}
B = \frac{\sqrt{m}}{1-\sqrt{m}}\,,~~A^2 =  \frac{4 \sqrt{m} \beta^2}
{(1-\sqrt{m})^2}\,.
\ee

{\bf Solution X} \\

The symmetric $\phi^4$ Eq. (\ref{phieq}) also admits another periodic 
solution \cite{ks4}
\be\label{4.51}
\phi(x) = \frac{A \sn(\beta x,m)}{1+B\cn^2(\beta x,m)}\,,~~B > 0\,,
\ee
provided
\bea\label{4.52}
&&0 < m < 1\,,~~B = \frac{\sqrt{m}}{1-\sqrt{m}}\,, \nonumber \\
&&a = [6\sqrt{m}-(1+m)]\beta^2\,,~~b A^2 = -8\sqrt{m} \beta^2\,.
\eea
On using the identity \cite{as}
\be\label{4.53}
\sn(y+\Delta,m)+\sn(y-\Delta,m) = \frac{2\sn(y,m) \frac{\cn(\Delta,m)}
{\dn(\Delta,m)}}{1+B\cn^2(y,m)}\,,~~B = \frac{m \sn^2(\Delta,m)}
{\dn^2(\Delta,m)}\,, 
\ee
one can re-express the solution (\ref{4.51}) as superposition of two 
periodic kink solutions
\be\label{4.54}
\phi(x) = i\sqrt{\frac{2m}{|b|}}\beta  
\bigg [\sn(\beta x +\Delta, m)+\sn(\beta x -\Delta, m) \bigg ]\,.
\ee
Here $\Delta$ is defined by $\sn(\sqrt{m}\Delta,1/m) = \pm m^{1/4}$,
where use has been made of the identity (\ref{4.47}).

It then follows that the nKdV-1 Eq. (\ref{3.4}) also admits the solution
($c < 0$)
\be\label{4.55}
u(\xi) = \frac{A \sn(\beta \xi,m)}{1+B\cn^2(\beta \xi,m)}\,,
~~B > 0\,,
\ee
provided
\bea\label{4.56}
&&0 < m < 1\,,~~B = \frac{\sqrt{m}}{1-\sqrt{m}}\,,
~~A^2 = 8|c|\sqrt{m} \beta^2\,.
\eea
In Fig. (7) we plot $u(\xi)$ of Eq. (\ref{4.55}) vs  $\xi$ in case $\beta = c = 1,
m = 1/2$.  
\begin{figure}
\includegraphics[width=0.4\linewidth]{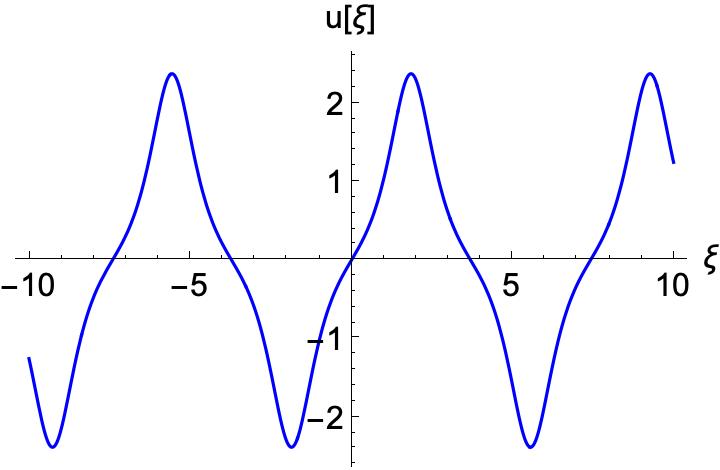}
\caption{The solution $u(\xi)$ of (\ref{4.55})  vs $\xi$ for $\beta = c = 1, 
m=1/2$. }
\label{sol10}
\end{figure}

It follows that the focusing nmKdV-1 Eq. (\ref{3.6}) also admits the 
solution (\ref{4.55}) provided
\bea\label{4.57}
&&0 < m < 1\,,~~B = \frac{\sqrt{m}}{1-\sqrt{m}}\,,
~~A^2 = 4\sqrt{m} \beta^2\,.
\eea

{\bf Solution XI} \\

The symmetric $\phi^4$ Eq. (\ref{phieq})  
admits another periodic solution \cite{ks4}
\be\label{4.58}
\phi(x) = \frac{A \sn(\beta x,m) \cn(\beta x,m)}{1+B\cn^2(\beta x,m)}\,,
\ee
provided
\bea\label{4.59}
&&0 < m < 1\,,~~B = \frac{1-\sqrt{1-m}}{\sqrt{1-m}}\,.
\nonumber \\
&&a = (2-m-6\sqrt{1-m})\beta^2\,,~~ b A^2 
= -\frac{8(1-\sqrt{1-m})^2 \beta^2}{\sqrt{1-m}}\,.
\eea 

On using the identity \cite{as}
\bea\label{4.60}
&&\dn(y-\Delta,m) -\dn(y+\Delta,m) = \frac{2m\sn(\Delta,m) \cn(\Delta,m) 
\sn(y,m) \cn(y,m)} {\dn^2(\Delta,m)[1+ B \cn^2(y)]}\,, \nonumber \\
&&B = \frac{m \sn^2(\Delta,m)}{\dn^2(\Delta,m)}\,. 
\eea
one can re-express solution (\ref{4.59}) as a superposition of two 
periodic pulse-like solutions, i.e.
\be\label{4.61}
\phi(x) = \beta \sqrt{\frac{2}{|b|}} \bigg (\dn[\beta x -\frac{K(m)}{2},m] 
- \dn[\beta x +\frac{K(m)}{2},m] \bigg )\,.
\ee

It then follows that the  nKdV-1 Eq. (\ref{3.4}) also admits the solution 
($c < 0$)
\be\label{4.62}
u(\xi) = \frac{A \sn(\beta x,m) \cn(\beta x,m)}{1+B\cn^2(\beta x,m)}\,,
\ee
provided
\bea\label{4.63}
&&0 < m < 1\,,~~B = \frac{1-\sqrt{1-m}}{\sqrt{1-m}}\,.
\nonumber \\
&&A^2 = |c|\frac{8(1-\sqrt{1-m})^2 \beta^2}{\sqrt{1-m}}\,.
\eea 
In Fig. (8) we plot $u(\xi)$ of Eq. (\ref{4.63}) vs  $\xi$ in case $\beta = c = 1,
m = 1/2$.  
\begin{figure}
	\includegraphics[width=0.4\linewidth]{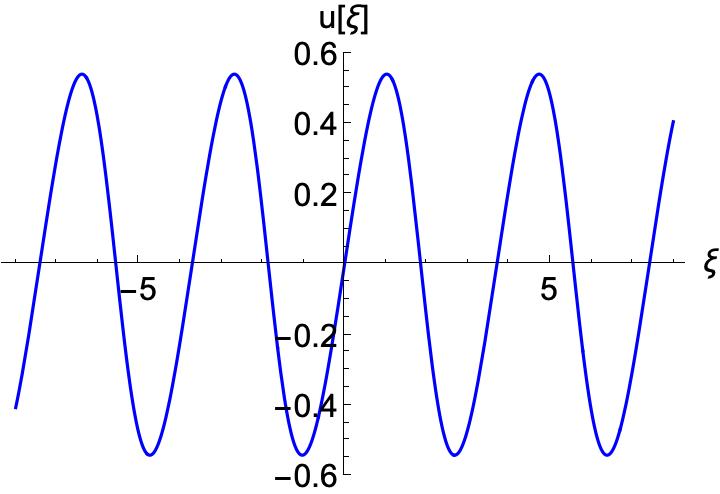}
	\caption{The solution $u(\xi)$ of Eq. (\ref{4.63})  $u(\xi)$  for $\beta=1$, $m=1/2$, $c=1$. 	}
	\label{sol11}
\end{figure}

It also follows that the focusing nmKdV-1 Eq. (\ref{3.6}) admits the solution 
(\ref{4.62}) provided
\bea\label{4.64}
&&0 < m < 1\,,~~B = \frac{1-\sqrt{1-m}}{\sqrt{1-m}}\,.
\nonumber \\
&&A^2 = \frac{4(1-\sqrt{1-m})^2 \beta^2}{\sqrt{1-m}}\,.
\eea 

{\bf Solution XII} \\

The symmetric $\phi^4$ Eq. (\ref{phieq}) also admits another periodic 
solution \cite{ks4}
\be\label{4.65}
\phi(x) = \frac{A \dn(\beta x,m)}{1+B\cn^2(\beta x,m)}\,,
\ee
provided
\bea\label{4.66}
&&0 < m < 1\,,~~B =  \frac{1-\sqrt{1-m}}{\sqrt{1-m}}\,,  
\nonumber \\
&&a = [2-m+6\sqrt{1-m}]\beta^2\,,~~b A^2 = -\frac{8}{\sqrt{1-m}} \beta^2\,.
\eea 

On using the identity \cite{as}
\be\label{4.67}
\dn(y+\Delta,m)+\dn(y-\Delta,m) = \frac{2\dn(y,m)}
{\dn(\Delta,m) [1+B\cn^2(y,m)]}\,,~~B = \frac{m \sn^2(\Delta,m)}
{\dn^2(\Delta,m)}\,,
\ee
one can re-express the solution (\ref{4.67}) as superposition of two periodic 
pulse-like solutions, i.e.
\be\label{4.68}
\phi(x) = \sqrt{\frac{2}{|b|}} \beta  
\bigg (\dn[\beta x +K(m)/2, m]+\dn[\beta x -K(m)/2, m] \bigg )\,.
\ee

It then follows that the nKdV-1 Eq. (\ref{3.4}) also admits the solution 
($c < 0$) 
\be\label{4.69}
u(\xi) = \frac{A \dn(\beta x,m)}{1+B\cn^2(\beta x,m)}\,,
\ee
provided
\bea\label{4.70}
&&0 < m < 1\,,~~B =  \frac{1-\sqrt{1-m}}{\sqrt{1-m}}\,. 
\nonumber \\
&&A^2 = \frac{8|c|}{\sqrt{1-m}} \beta^2\,.
\eea 

It also follows that the focusing nmKdV-1 Eq. (\ref{3.6}) admits the 
solution (\ref{4.69}) provided 
\bea\label{4.71}
&&0 < m < 1\,,~~B = \frac{1-\sqrt{1-m}}{\sqrt{1-m}}\,, 
\nonumber \\
&&\alpha = A^2 = \frac{4}{\sqrt{1-m}} \beta^2\,.
\eea

%
%This work raises several open questions, some of which are
\section{Symmetries and special solutions of the  nKdV equation} 
% \subsection{Symmetries of the Lou Form of the  equation}
In this section we first consider the symmetries of the nKdV equation 
(\ref{3.8}) and then obtain the rational solutions of the nKdV equation.

\subsection{Symmetries of the nKdV Eq. (\ref{3.8})}

Let us now consider the various symmetries of the nKdV Eq. (\ref{eq8}).
Once we know its symmetries, we can immediately obtain the symmetries of $w, v$, or $\tw,v$.
%Note that $w, v$ satisfy the coupled  Eq. (\ref{eq7}).

\begin{enumerate}

\item   The nKdV Eq. (\ref{eq8})  is invariant under $u \rightarrow -u$. Thus if 
$u_1$ is a solution of  Eq. (\ref{eq8}) then so is $-u_1$. Under this
transformation both $w, v$ are invariant.

\item  Under $u \rightarrow \pm iu$   Eq. (\ref{eq8})  goes over to 
\be\label{6.1} 
u(x,t) u_{xxt} -u_{xx} u_t + 2 u^3 u_x = 0\,,
\ee
thereby implying that if $u_1$ is a solution then $\pm i u_1$ is also a 
solution. Under this transformation while $v$ remains unchanged, $w$ goes
to $-w$.
% Using Eq. (\ref{2}), it implies
%that corresponding to a solution $u_1$ of Eq. (\ref{uu}),
%This implies that  if $w_1,  v_1$ are 
%the solutions of Eq. (\ref{u3}), then corresponding to the Eq. (\ref{4}), the 
%corresponding solutions are $-w_1, v_1$.

\item   Eq. (\ref{eq8})  is invariant under
\be\label{6.2}
x \rightarrow \alpha x\,,~~ t \rightarrow t\,,~~u \rightarrow \alpha^{-1/2} u\, .
\ee
Under this transformation, while $w \rightarrow \alpha^{-1} w, 
v \rightarrow \alpha^{-2} v$\,.

\item  Eq. (\ref{eq8})  is invariant under
\be\label{6.3}
x \rightarrow x\,,~~ t \rightarrow \alpha t\,,~~u \rightarrow \alpha^{-1/2} u\,.
\ee
Under this transformation, while $w \rightarrow \alpha^{-1} w, 
v \rightarrow v$\,.

\item   Eq. (\ref{eq8})  is invariant  under
\be\label{6.4}
x \rightarrow \alpha x\,,~~ t \rightarrow \alpha^{-1} t\,,
~~u \rightarrow \alpha^{-1/2} u\, .
\ee
Under this transformation, while $w \rightarrow \alpha^{-1} w, 
v \rightarrow \alpha^{-2} v$\,.

\item   Eq. (\ref{eq8}) is invariant under 
\be\label{6.5}
x \rightarrow \alpha x, t \rightarrow \alpha t, u \rightarrow \alpha^{-1}\,.
\ee
Note that under this transformation $w \rightarrow \alpha^{-2} w$ and
$v \rightarrow \alpha^{-2} v$.

\end{enumerate}

\subsection{ Rational Solutions of nKdV Eq. (\ref{eq8}) }

If we make the ansatz
\be\label{6.6}
u(x,t) = \frac{1}{t} f(\eta)\,,~~\eta = \frac{x}{t}\,,
\ee
then  Eq. (\ref{eq8})  takes the form
\be\label{6.7}
\eta f f''' + (2f - \eta f')f'' - 2 f^3 f' = 0 \,. 
\ee
Eq. (\ref{6.7}) can be immediately integrated yielding
\be\label{6.8}
\eta f f''+f f' - \eta (f')^2 -(1/2) f^4 = c\,,
\ee
where $c$ is a constant of integration. If we solve 
 Eq. (\ref{6.8}) and obtain $f$ we then find $u$ from Eq. (\ref{6.6}).
Eq. (\ref{6.8}) is very close to a special case of the Painleve-III \cite{as}.  
The most general form of Painleve-III is 
\be\label{6.9}
\eta f f" + f f' -\eta (f')^2 -\delta \eta - \beta f - \alpha f^3 
-\gamma \eta f^4 = 0\,.
\ee
On comparing Eqs. (\ref{6.8}) and (\ref{6.9}) we see that in the limit 
$\delta = \beta = \gamma = 0$, both equations are identical but for
the $f^4$ term in Eq. (\ref{6.8}) and $f^3$ term in Eq. (\ref{6.9}).
We now present a solution of Eq. (\ref{6.8})
\be\label{6.10}
f(\eta) = \frac{i\sqrt{2\alpha}}{\eta+\alpha} ,    
\ee  
where $\alpha$ is any real number. The corresponding $u(x,t)$ is
\be\label{6.11}
u(x,t) = \frac{i\sqrt{2\alpha}}{x+\alpha t} \,. 
\ee
This is a singular solution. A related nonsingular solution is 
\be\label{6.12}
u(x,t) = \frac{i\sqrt{2\alpha}}{x+\alpha t + i \beta} ,  
\ee
where $\alpha, \beta$ are arbitrary real numbers.
Using Eq. (\ref{eq9}), the corresponding $v, w$ are easily obtained.

\section{Symmetries and Novel solutions of the nmKdV-1 and nmKdV-2 equations}

We will first discuss the symmetries of the nmKdV-1 and the nmKdV-2 equations, 
both focusing and defocusing. We will then obtain the rational solutions of the 
focusing as well as the defocusing nmKdV-1 as well as nmKdV-2. Finally, we
will show that for both the nmKdV-1 as well as the nmKdV-2 equations we can 
decouple $x$ and $t$ and obtain novel solutions for both the focusing and the 
defocusing cases. Remarkably, it turns out that not only all the symmetries of
the nmKdV-1 and the nmKdV-2 are the same but even the rational and the novel 
solutions (where $x$ and $t$ decouple) that we have found are the same and are 
valid under the same conditions.
\subsection{Symmetries of nmKdV-I and nmKdV-2 equations}

We will observe that the symmetries of the nmKdV-1 as well as of the 
nmKdV-2 equations are identical. 

\begin{enumerate}

\item The focusing and the defocusing nmKdV-1 and nmKdV-2  Eqs. (\ref{10}) 
to (\ref{13}) are invariant under $u \rightarrow -u$. Thus if $u_1$ is a 
solution of these equations, then $-u_1$ will also be solution of these
equations.

\item The focusing and the defocusing nmKdV-1 and nmKdV-2  equations, 
Eqs. (\ref{10}) to (\ref{13}), are invariant under $ \rightarrow -x$. 

\item Under $u \rightarrow \pm iu$, the focusing (defocusing) nmKdV-1 
Eq. (\ref{10}) [(Eq. (\ref{11})] goes over to the defocusing (focusing) nmKdV-1
Eq. (\ref{11}) [Eq. (\ref{10})].  
		
\item Under $u \rightarrow \pm iu$, the focusing (defocusing) nmKdV-2 
Eq. (\ref{12}) [(Eq. (\ref{13})] goes over to the defocusing (focusing) nmKdV-1
Eq. (\ref{13}) [Eq. (\ref{12})].  
		
\item Under $x \rightarrow \pm ix$, the focusing (defocusing) nmKdV-1 
Eq. (\ref{10}) [(Eq. (\ref{11})] goes over to the defocusing (focusing) nmKdV-1
Eq. (\ref{11}) [Eq. (\ref{10})].  
		
\item Under $x \rightarrow \pm ix$, the focusing (defocusing) nmKdV-2 
Eq. (\ref{12}) [(Eq. (\ref{13})] goes over to the defocusing (focusing) nmKdV-1
Eq. (\ref{13}) [Eq. (\ref{12})].  
		
\item The focusing as well as defocusing nmKdV-1 as well as nmKdV-2, Eqs. 
(\ref{10}) to (\ref{13}), are invariant under
\be\label{6.13}
x \rightarrow \alpha x\,,~~u \rightarrow \alpha^{-1} u\,.
\ee

\item The focusing as well as the defocusing nmKdV-1 as well as nmKdV-2, Eqs. 
(\ref{10}) to (\ref{13}), are invariant under
\be\label{6.14}
t \rightarrow \alpha t\,.
\ee

\end{enumerate}

\subsection{Rational Solutions of the nmKdV-1 }

As in the  nKdV case, if we make the ansatz
\be\label{6.15}
u(x,t) = \frac{1}{t} f(\eta)\,,~~\eta = \frac{x}{t}\,,
\ee
then focusing nmKdV-1 Eq. (\ref{10}) takes the form
\be\label{6.16}
\eta f f''' + (3f - \eta f')f'' -2(f')^2 + 4\eta f^3 f'+4 f^4 = 0\,,
\ee
while the defocusing nmKdV-1 Eq.(\ref{11}) takes the form
\be\label{6.17}
\eta f f''' + (3f - \eta f')f'' -2(f')^2 - 4\eta f^3 f'- 4 f^4 = 0\,.
\ee

For the focusing case we find that 
\be\label{6.18}
f(\eta) =\frac{A} {\eta+\alpha}\,, 
\ee
with  $\alpha$ any real number is an exact solution of Eq. (\ref{6.16}) 
provided $ A = \pm i$. Thus the corresponding singular $ u(x,t)$ is
\be\label{6.19}
u(x,t) = \pm   \frac{i} {x+\alpha t}\,.   
\ee
If instead we choose $\alpha$ to be pure imaginary then we obtain complex
but nonsingular solution of focusing nmKdV-1
\be\label{6.20}
u(x,t) = \pm   \frac{i} {x+i\alpha t}\,,   
\ee
where $\alpha$ is any real number. Yet another nonsingular but complex solution
of focusing nmKdV-1 Eq. (\ref{10}) is
\be\label{6.21}
u(x,t) = \pm \frac{i}{ x+\alpha t + i \beta} \,, 
\ee
with $\alpha, \beta$ being any arbitrary real numbers.

For the defocusing nmKdV Eq. (\ref{6.17}) one instead finds that
\be\label{6.22}
f(\eta) = \frac{A} {\eta+\alpha}\,,   
\ee
with $\alpha$ being any real number,
is an exact solution of Eq. (\ref{6.17}) provided $A = \pm 1$. The 
corresponding singular $u(x,t)$ is
\be\label{6.23}
u(x,t) = \pm \frac{1}{ x+\alpha t}\,. 
\ee 
If instead we choose $\alpha$ to be pure imaginary then we obtain a complex
but nonsingular solution of focusing nmKdV-2
\be\label{6.24}
u(x,t) = \pm   \frac{1} {x+i\alpha t}\,,   
\ee
where $\alpha$ is any real number. Yet another nonsingular but complex solution
of the defocusing nmKdV-1 is
\be\label{6.25}
u(x,t) = \pm \frac{1}{ x+\alpha t + i \beta}\,, 
\ee
with $\alpha,\beta$ being  arbitrary real numbers.

\subsection{Unusual solutions of nmKdV-1}

We now show that the focusing and defocusing nmKdV-1 Eqs. (\ref{10}) and 
(\ref{11}) admit solutions in which $t$ and $x$ decouple. In particular, if we 
start with the ansatz 
\be\label{6.26}
u(x,t) = \theta(t) g(x)\,,
\ee
where $\theta(t) = 1$ $(-1)$ for $t > (< )$ $0$, then it is easy to show that 
Eq. (\ref{6.26}) is an exact solution of the nmKdV-1 Eq. (\ref{10}) provided
\be\label{6.27}
g g_{xx} -g_{x}^{2} + 4g^4 = 0\,.
\ee
%There must be several
%solutions of nonlinear Eq.(\ref{7.2}). Unfortunately, so far I have been
%able to obtain one solution. In particular
A particular solution of  Eq. (\ref{6.27})  is 
\be\label{6.28a}
g(x) = A \sech(\beta x)\,,
\ee
provided
\be\label{6.29a}
2A = \beta\,.
\ee
Another solution is 
\be\label{6.30a}
g(x) = \pm \frac{1}{2(\alpha+ i x)}\,, 
\ee
where $\alpha$ is an arbitrary real number.

If we use the same ansatz (\ref{6.28a}) and use it in the defocusing 
Eq. (\ref{11}), we instead find that it is an exact solution of Eq. (\ref{10}) 
provided $g(x)$ satisfies
\be\label{6.31a}
g g_{xx} -g_{x}^{2} - 4g^4 = 0\,.
\ee
Two solutions of this equation are
\be\label{6.28}
g(x) = \pm iA \sech(\beta x)\,,
\ee
provided
\be\label{6.29}
2A = \beta\,, 
\ee
\be\label{6.30}
g(x) = \pm \frac{i}{2(x+ i \alpha)} .
\ee

\subsection{Rational Solutions of nmKdV-2 }

As in the nmKdV-1 case, if we make the ansatz
\be\label{6.31}
u(x,t) = \frac{1}{t} f(\eta)\,,~~\eta = \frac{x}{t}\,,
\ee
then the focusing nmKdV-2 Eq. (\ref{12}) takes the form
\bea\label{6.32}
&&4 f' f''' + \eta f' f'''' +20 f^2 (f')^2 - 8\eta f^2 f' f'' 
+12\eta f (f')^3  \nonumber \\
&&-3(f'')^2-\eta f'' f''' - 4f^3 f'' = 0\,,
\eea
while the defocusing nmKdV-2 Eq. (\ref{13}) takes the form
\bea\label{6.33}
&&4 f' f''' + \eta f' f'''' -20 f^2 (f')^2 + 8\eta f^2 f' f'' 
-12\eta f (f')^3  \nonumber \\
&&-3(f'')^2-\eta f'' f''' + 4f^3 f'' = 0\,. 
\eea

One finds  that Eq. (\ref{6.32}) has the solution
\be\label{6.34}
f(\eta) = \frac{A}{\eta+\alpha}\,,
\ee
with  $\alpha$ any real number
provided $ A = \pm i$. Thus the corresponding singular $u(x,t)$ is
\be\label{6.35}
u(x,t) = \pm \frac{i} {x+\alpha t}\,.   
\ee
If we instead choose $\alpha$ to be pure imaginary, then we obtain
the nonsingular but complex solution of focusing nmKdV-2 
\be\label{6.36}
u(x,t) = \pm \frac{i}{x+i\alpha t}\,,
\ee
where $\alpha$ is any real number. One can show that another complex
but nonsingular solution of focusing nmKdV-2 Eq. (\ref{12}) is
\be\label{6.37}
u(x,t) = \pm \frac{i}{ x+\alpha t + i \beta} \,, 
\ee
with $\alpha, \beta$ being any arbitrary real numbers.

For the defocusing nmKdV Eq. (\ref{6.33}) one instead finds that
\be\label{6.38}
f(\eta) = \frac{A} {\eta+\alpha}\,,   
\ee
with $\alpha$ being any real number,
is an exact solution of Eq. (\ref{6.33}) provided $A = \pm 1$. The 
corresponding singular $u(x,t)$ is
\be\label{6.39}
u(x,t) = \pm \frac{1}{ x+\alpha t}\,.
\ee 
Instead if we choose $\alpha$ to be pure imaginary, then we have a nonsingular
but complex solution of $f$ and hence $u$ is given by 
\be\label{6.40}
u(x,t) = \pm \frac{1}{x+i\alpha t}\,,
\ee
where $\alpha$ is any real number. Another 
nonsingular but complex solution of defocusing nmKdV-2 is
\be\label{6.41}
u(x,t) = \pm \frac{1}{ x+\alpha t + i \beta}\,, 
\ee
with $\alpha,\beta$ being  arbitrary real numbers.

It is amusing to note that even though the nmKdV-1 and nmKdV-2 equations are
very different and yet their focusing as well as defocusing rational solutions 
are identical and are satisfied under the same conditions.

\subsubsection{\bf Unusual solutions of nmKdV-2}

We now show that the nmKdV-2 Eqs. (\ref{12}) and (\ref{13}) admit solutions 
in which $t$ and $x$ decouple. In particular, if like in nmKdV-1, we start with
the ansatz 
\be\label{6.42}
u(x,t) = \theta(t) g(x)\,,
\ee
then it is easy to show that Eq. (\ref{6.42}) is an exact solution of the
focusing nmKdV-2 Eq. (\ref{12}) provided
\be\label{6.43}
g_{x} g_{xxx} -g_{xx}^{2} + 16 g^2 (g_{x})^2 - 4g^3 g_{xx} = 0\,.
\ee
%There must be several
%solutions of nonlinear Eq.(\ref{7.2}). Unfortunately, so far I have been
%able to obtain one solution. In particular
A particular solution of  Eq. (\ref{6.43})  is 
\be\label{6.44}
g(x) = A \sech(\beta x)\,,
\ee
provided
\be\label{6.45}
2A = \beta\,.
\ee
Another solution of Eq. (\ref{6.43}) is 
\be\label{6.46}
g(x) = \pm \frac{1}{2(\alpha+ i x)}\,, 
\ee
where $\alpha$ is an arbitrary real number.

If we use the same ansatz (\ref{6.42}) and use it in Eq. (\ref{13}), we
instead find that it is an exact solution provided $g(x)$ satisfies
\be\label{6.47}
g_{x} g_{xxx} -g_{xx}^{2} - 16 g^2 (g_{x})^2 + 4g^3 g_{xx} = 0\,.
\ee
One of the two solutions of this equation is 
\be\label{6.48}
g(x) = \pm iA \sech(\beta x)\,,
\ee
provided
\be\label{6.49}
2A = \beta\,.
\ee
Another solution of Eq. (\ref{6.47}) is
\be\label{3.39}
g(x) = \pm \frac{i}{2(x+ i \alpha)} \,. 
\ee

It is amusing to note that even though the nmKdV-1 and nmKdV-2 equations are
very different and yet their rational, as well as the novel solutions (where 
$x$ and $t$ decouple) are not only the same but are satisfied under the same 
conditions.

\subsection{Traveling Wave Solutions of nmKdV-2}

In view of the fact that the rational as well as the novel solutions (where $x$
and $t$ decouple) that we have obtained are the same for the nmKdV-1 and the 
nmKdV-2 equations, it is natural to enquire if the nmKdV-2 equation also admits
the same traveling wave solutions as admitted by the nmKdV-1 equation and 
under the same conditions. In order to get the traveling wave solutions of the 
nmKdV-2 equation, as in Sec. IV above, we introduce the variable $\xi = x-ct$. In 
terms of this variable the focusing and the defocusing Eqs. (\ref{12}) and 
(\ref{13}) respectively, take the form
\be\label{3.40}
u_{\xi}u_{\xi\xi\xi\xi} -u_{\xi\xi} u_{\xi\xi\xi} +12 u u_{\xi}^3 = 0\,,
\ee
\be\label{3.41}
u_{\xi}u_{\xi\xi\xi\xi} -u_{\xi\xi} u_{\xi\xi\xi} -12 u u_{\xi}^3 = 0\,.
\ee
We have explicitly checked that the focusing nmKdV-1 solutions I, III of 
Sec. IV as well as the solutions XIIIA, XIVA, XVIA, XVIIA, XXA, XXIIIA of the 
Appendix are also the solutions of the focusing nmKdV-2 Eq. (\ref{3.40}) and 
under the same conditions as applicable to the corresponding nmKdV-1 solutions.
Similarly, we have also explicitly checked that the defocusing nmKdV-1 
solutions II, IV of Sec. IV as well as the solutions XVA, XVIIIA, XXIA, XXIIA  
of the Appendix are also the solutions of the defocusing nmKdV-2 Eq. (\ref{3.41}) 
and under the same conditions as applicable to the corresponding nmKdV-1 solutions. 
We suspect that the remaining 18 traveling wave solutions of the nmKdV-1 equation are 
also the solutions of the corresponding nmKdV-2 equation and under the same 
conditions as for the nmKdV-1 solutions.

\section{Open Problems}

This work raises several open questions, some of which are:

\begin{enumerate} 

\item In Sec. IV and in the Appendix, we have obtained 32 moving solutions of 
both the nKdV-1 and the nmKdV-1 equations by making use of the known static solutions 
of the symmetric $\phi^4$ equation. The obvious question is, has one exhausted all
the static solutions of $\phi^4$ equation and hence of nKdV-1 and nmKdV-1?

\item In Sec. VI we have shown that for both the nmKdV-1 and the nmKdV-2 and 
for both the focusing and the defocusing cases, one can choose a novel ansatz so 
that $t$ and $x$ decouple and one obtains nonlinear equations in $x$. 
While we have obtained few solutions of these nonlinear equations, clearly 
we have not obtained all the possible solutions of these equations. It will be 
interesting to find at least few more such solutions. 

\item In Sec. VI for both the nmKdV-1 and the nmKdV-2 equations and for both the 
focusing and the defocusing cases, we have obtained few rational solutions. Clearly 
we have not obtained all the possible rational solutions of these equations. It
will be interesting to find at least few more rational solutions. 

\item Even though the nmKdV-1 and nmKdV-2 equations are very different, 
we found that the rational solutions as well as the novel solutions where $x$
and $t$ decouple are identical for both focusing and defocusing 
nmkdV-1 and nmKdV-2 equations. Besides, we have also explicitly checked that 
this is also true for the 14 moving traveling wave solutions. The first 
question is whether this is also true for the remaining 18 traveling wave 
solutions of the nmKdV-1 that we have obtained in Sec. IV and in the Appendix? 
The next more general question is whether this is true in general for all the solutions 
of the nmKdV-1 and nmKdV-2 equations or whether we can find few different solutions 
of these equations. 

\item The  KdV equation is an integrable equation which can be 
solved through inverse scattering and the corresponding linear problem is
essentially a Schr\"odinger-like equation. Since the nKdV equation is also
integrable, the obvious question is if it too can be solved using the 
inverse scattering method. If so, what is the corresponding linear problem? Is 
it again a  Schr\"odinger-like equation?

\item The one-soliton solution of the KdV equation for fixed time is 
essentially a reflectionless potential in the context of the corresponding 
Schr\"odinger-like equation. What about the corresponding one-soliton 
solutions of the nKdV equation? For fixed time, does it again correspond to 
a reflectionless potential?

\item The periodic one-soliton solution of the KdV equation for fixed time is 
known to have a novel band structure. In particular, there is a band, a band gap
followed by a continuum. What about the corresponding periodic one-soliton 
solution of the nKdV equation? For fixed time, does it have a novel band 
structure? 

\end{enumerate}

We hope to address some of these questions in the near future.

\section{Acknowledgment}

One of us, AK. is grateful to Indian National Science Academy (INSA) for the 
award of INSA Honorary Scientist Position at Savitribai Phule Pune 
University. The work at LANL was carried out under the auspices of the US 
Department of Energy NNSA under Contract No. 89233218CNA000001.

\section{Appendix}
In Sec. IV we have presented 12 traveling wave solutions of the nKdV-1 and the
nmKdV-1 equations by comparing them with the corresponding known static solutions 
of the symmetric $\phi^4$ equation. We now present the remaining 20 solutions.

{\bf Solution XIIIA}

One of the well known periodic pulse solutions of Eq. (\ref{phieq}) 
is
\be\label{A1}
\phi(x) = A \dn(\beta x)\,,
\ee
provided
\be\label{A2}
b A^2 = -2\beta^2\,,~~a = (2-m)\beta^2\,.
\ee

Hence the corresponding periodic traveling wave solution of the nKdV-1  Eq. 
(\ref{3.4}) is ($c < 0$)
\be\label{A3} 
u(\xi) = \sqrt{2|c|} \beta \dn(\beta \xi,m)\,,~~\xi = x-ct\,.
\ee

On the other hand, the corresponding traveling wave solution of the focusing 
nmKdV-1 Eq. (\ref{3.6}) is 
\be\label{A4}
u(\xi) = \beta \dn(\beta \xi,m)\,.
\ee

{\bf Solution XIVA}

Eq. (\ref{phieq}) also admits a complex PT-invariant periodic pulse solution 
with $PT$ eigenvalue +$1$ \cite{ks1}
\be\label{A5}
\phi(x) = A \dn(\beta x, m) +i B \sqrt{m} \sn(\beta x, m)\,,
\ee
provided
\be\label{A6}
B = \pm A\,,~~2bA^2 = -\beta^2\,,~~a = -\frac{(2m-1)\beta^2}{2}\,.
\ee
In the limit $m = 1$  Eq. (\ref{A5} )  goes over to the complex
hyperbolic pulse solution with PT-eigenvalue +$1$
\be\label{5.24}
\phi(x) = A \sech(\beta x) +i B \tanh(\beta x)\,.
\ee

It then follows that the nKdV-1 Eq. (\ref{3.4}) has the solution  ( $c<0$) 
\be
u(\xi)  =\sqrt{\frac{|c|}{2}} \beta [(\dn(\beta \xi, m) 
\pm i\sqrt{m} \sn(\beta \xi, m)]\,.
\ee

On the other hand, the focusing nmKdV-1 Eq. (\ref{3.6})  has  the solution 
\be\label{A7}
u(\xi)  = \frac{\beta}{2} [\dn(\beta \xi, m) \pm i\sqrt{m} \sn(\beta \xi, m)]\,. 
\ee

{\bf Solution XVA}

Remarkably, Eq. (\ref{phieq}) also admits a complex PT-invariant periodic kink
solution with $PT$ eigenvalue $-1$
\be\label{A8}
\phi(x) = A \sqrt{m} \sn(\beta x, m) +i B \dn(\beta x, m)\,,
\ee
provided
\be\label{A9}
B = \pm A\,,~~2bA^2 =  \beta^2\,,~~a = -\frac{(2m-1)\beta^2}{2}\,.
\ee

It then follows that the  nKdV-1 Eq. (\ref{3.4}) also admits the solution 
($c > 0$) 
\be\label{A10}
u(\xi) = \sqrt{\frac{c}{2}}\beta[\sqrt{m} \sn(\beta \xi, m) 
\pm i\dn(\beta \xi, m)]\,. 
\ee

It also follows that the defocusing nmKdV-1 Eq. (\ref{3.8}) admits the solution 
\be\label{A11}
u(\xi) = \frac{\beta}{2} [\sqrt{m} \sn(\beta \xi, m) 
\pm i\dn(\beta \xi, m)]\,. 
\ee

{\bf Solution XVIA}

Eq. (\ref{phieq}) also admits the periodic pulse solution
\be\label{A12}
\phi(x) = \frac{A\sqrt{1-m}}{\dn(\beta x, m)}\,,
\ee
provided
\be\label{A13}
0 < m < 1\,,~~bA^2  = -2\beta^2\,,~~a = (2-m)\beta^2\,.
\ee

It then follows that  nKdV-1 Eq. (\ref{3.4}) also admits the solution 
($c < 0$)
\be\label{A14}
u(\xi) = \frac{\sqrt{2(1-m)|c|\beta}}{\dn(\beta \xi,m)}\,,~~0 < m < 1\,.
\ee

It also follows that the focusing nmKdV-1 Eq. (\ref{3.6}) also admits the 
solution 
\be\label{A15}
u(\xi) = \frac{\sqrt{(1-m)\beta}}{\dn(\beta \xi,m)}\,,~~0 < m < 1\,.
\ee

{\bf Solution XVIIA}

Eq. (\ref{phieq}) in fact also admits the periodic kink solution
\be\label{A16}
\phi(x) = \frac{A\sqrt{m(1-m)}\sn(\beta x,m)}{\dn(\beta x, m)}\,,
\ee
provided
\be\label{A17}
0 < m < 1\,,~~bA^2  = -2\beta^2\,,~~a = (2m-1)\beta^2\,.
\ee

It then follows that the nKdV-1  Eq. (\ref{3.4}) also admits the solution 
($c < 0$)
\be\label{A18}
u(\xi) = \frac{\sqrt{2|c| m(1-m)} \beta \sn(\beta \xi,m)}{\dn(\beta \xi, m)}\,,
~~0 < m < 1\,.
\ee

It also follows that the focusing nmKdV-1 Eq. (\ref{3.6}) admits the periodic 
kink solution 
\be\label{A19}
u(\xi) = \frac{\sqrt{m(1-m)}\beta \sn(\beta x,m)}{\dn(\beta x, m)}\,,
~~0 < m < 1\,.
\ee

{\bf Solution XVIIIA}

Eq. (\ref{phieq}) also admits the periodic pulse solution
\be\label{A20}
\phi(x) = \frac{A\sqrt{m}\cn(\beta x,m)}{\dn(\beta x, m)}\,,
\ee
provided
\be\label{A21}
0 < m < 1\,,~~ bA^2  = 2\beta^2\,,~~a = -(1+m)\beta^2\,.
\ee

It then follows that nKdV-1  Eq. (\ref{3.4}) admits the solution 
($c > 0$)
\be\label{A22}
u(\xi) = \frac{\sqrt{2m c} \beta \cn(\beta \xi,m)}{\dn(\beta \xi, m)}\,,
~~0 < m < 1\,.
\ee

It also follows that the defocusing nmKdV-1 Eq. (\ref{3.8}) admits the solution 
\be\label{A23}
u(\xi) = \frac{\sqrt{m} \beta \cn(\beta \xi,m)}{\dn(\beta \xi, m)}\,,~~0 < m < 1\,.
\ee

{\bf Solution XIXA}

Remarkably, Eq. (\ref{phieq}) also admits another periodic superposed solution
\be\label{A24}
\phi(x) = \frac{A\sqrt{1-m}}{\dn(\beta x, m)} 
+\frac{B\sqrt{m(1-m)}\sn(\beta x,m)}{\dn(\beta x, m)}\,,
\ee
provided
\be\label{A25}
0 < m < 1\,,~~B = \pm A\,,~~ 2bA^2  = -\beta^2\,,~~a = \frac{(1+m)\beta^2}{2}\,.
\ee

It then follows that  the nKdV-1 Eq. (\ref{3.4}) also admits the solution 
($c < 0\,, 0 < m < 1$)
\be\label{A26}
u(\xi) = \frac{\sqrt{|c|(1-m)}\beta}{\sqrt{2}} [{\dn(\beta \xi, m)} 
\pm \frac{\sqrt{m}\sn(\beta x,m)}{\dn(\beta \xi, m)}]\,. 
\ee

It also follows that the focusing nmKdV-1 Eq. (\ref{3.6}) admits the solution 
($0 < m < 1$)
\be\label{A27}
u(\xi) = \frac{\sqrt{(1-m)}\beta}{2} [{\dn(\beta \xi, m)} 
\pm \frac{\sqrt{m}\sn(\beta x,m)}{\dn(\beta \xi, m)}]\,. 
\ee

{\bf Solution XXA}

Eq. (\ref{phieq}) also admits a complex (but {\it not} PT-invariant) periodic 
superposed solution
\be\label{A28}
\phi(x) = \frac{A\sqrt{1-m}}{\dn(\beta x, m)} 
+\frac{iB\sqrt{m}\cn(\beta x,m)}{\dn(\beta x, m)}\,,
\ee
provided
\be\label{A29}
0 < m < 1\,,~~B = \pm A\,,~~ 2bA^2  = -\beta^2\,,
~~a = -\frac{(2m-1)\beta^2}{2}\,.
\ee

It then follows that  the nKdV-1 Eq. (\ref{3.4}) also admits the solution
($c < 0\,, 0 < m < 1$)
\be\label{A30}
u(\xi) = \frac{\sqrt{|c|}}{2}\beta [\frac{\sqrt{1-m}}{\dn(\beta \xi, m)} 
\pm \frac{i\sqrt{m}\cn(\beta \xi,m)}{\dn(\beta \xi, m)}]\,. 
\ee

It also follows that the focusing nmKdV-1 Eq. (\ref{3.6}) admits the solution 
($0 < m < 1$)
\be\label{A31}
u(\xi) = \frac{\beta}{2} [\frac{\sqrt{1-m}}{\dn(\beta \xi, m)} 
\pm \frac{i\sqrt{m}\cn(\beta \xi,m)}{\dn(\beta \xi, m)}]\,. 
\ee

{\bf Solution XXIA}

Eq. (\ref{phieq}) also admits a complex (but {\it not} PT-invariant) periodic 
superposed solution
\be\label{A32}
\phi(x) = \frac{A\sqrt{m}\cn(\beta x,m)}{\dn(\beta x, m)} 
+\frac{iB\sqrt{1-m}}{\dn(\beta x, m)}\,,
\ee
provided
\be\label{A33}
0 < m < 1\,,~~B = \pm A\,,~~ 2bA^2  = \beta^2\,,
~~a = -\frac{(2m-1)\beta^2}{2}\,.
\ee

It then follows that  the nKdV-1 Eq. (\ref{3.4}) also admits the solution 
($0 < m < 1, c > 0$)
\be\label{A34}
u(\xi) = \frac{\sqrt{c}\beta}{\sqrt{2}}[\frac{\sqrt{m}\cn(\beta \xi,m)}
{\dn(\beta \xi, m)} \pm i\frac{\sqrt{1-m}}{\dn(\beta \xi, m)}]\,. 
\ee

It also follows that the defocusing nmKdV-1 Eq. (\ref{3.8}) admits the solution
\be\label{A35}
u(\xi) = \frac{\beta}{2}[\frac{\sqrt{m}\cn(\beta \xi,m)}
{\dn(\beta \xi, m)} \pm i\frac{\sqrt{1-m}}{\dn(\beta \xi, m)}]\,. 
\ee

{\bf Solution XXIIA}

Eq. (\ref{phieq}) also admits a complex PT-invariant periodic 
superposed solution with PT-eigenvalue +$1$
\be\label{A36}
\phi(x) = \frac{A\sqrt{m}\cn(\beta x,m)}{\dn(\beta x, m)} 
+\frac{iB\sqrt{m(1-m})\sn(\beta x,m)}{\dn(\beta x, m)}\,,
\ee
provided
\be\label{A37}
0 < m < 1\,,~~B = \pm A\,,~~ 2bA^2  = \beta^2\,,
~~a = -\frac{(2-m)\beta^2}{2}\,.
\ee

It then follows that the nKdV-1 Eq. (\ref{3.4}) also admits the solution 
($0 < m < 1, c > 0$)
\be\label{A38}
u(\xi) = \frac{\sqrt{c m}\beta}{\sqrt{2}}[\frac{\cn(\beta \xi,m)}
{\dn(\beta \xi, m)} \pm i\frac{\sqrt{1-m}\sn(\beta \xi,m)}
{\dn(\beta \xi, m)}]\,. 
\ee

It also follows that the defocusing nmKdV-1 Eq. (\ref{3.8}) admits the solution
\be\label{A39}
u(\xi) = \frac{\sqrt{m}\beta}{2}[\frac{\cn(\beta \xi,m)}
{\dn(\beta \xi, m)} \pm i\frac{\sqrt{1-m}\sn(\beta \xi,m)}
{\dn(\beta \xi, m)}]\,. 
\ee

{\bf Solution XXIIIA}

Eq. (\ref{phieq}) also admits a complex PT-invariant periodic 
superposed solution with PT-eigenvalue $-1$
\be\label{A40}
\phi(x) = \frac{A\sqrt{m(1-m})\sn(\beta x,m)}{\dn(\beta x, m)}
+i\frac{B\sqrt{m}\cn(\beta x,m)}{\dn(\beta x, m)} 
\ee
provided
\be\label{A41}
0 < m < 1\,,~~B = \pm A\,,~~ 2bA^2  = -\beta^2\,. 
~~a = -\frac{(2-m)\beta^2}{2}\,.
\ee

It then follows that  the nKdV-1 Eq. (\ref{3.4}) also admits the solution 
($0 < m < 1, c < 0$)
\be\label{A42}
u(\xi) = \frac{\sqrt{c m}\beta}{\sqrt{2}}[\frac{\sqrt{1-m}\sn(\beta \xi,m)}
{\dn(\beta \xi, m)}] \pm i\frac{\cn(\beta \xi,m)}{\dn(\beta \xi, m)}] \,. 
\ee

It also follows that the focusing nmKdV-1 Eq. (\ref{3.6}) admits the solution
\be\label{A43}
u(\xi) = \frac{\sqrt{m}\beta}{2}[\frac{\sqrt{1-m}\sn(\beta \xi,m)}
{\dn(\beta \xi, m)}] \pm i\frac{\cn(\beta \xi,m)}{\dn(\beta \xi, m)}] \,. 
\ee

{\bf Solution XXIVA}

Eq. (\ref{phieq}) also admits a novel superposed periodic 
solution 
\be\label{A44}
\phi(x) = A \dn(\beta x,m) +\frac{B\sqrt{m(1-m})\sn(\beta x,m)}
{\dn(\beta x, m)}\,,
\ee
provided
\be\label{A45}
0 < m < 1\,,~~B = \pm A\,,~~ 2bA^2  = -\beta^2\,,
~~a = (1+m)\beta^2\,.
\ee

It then follows that  the nKdV-1 Eq. (\ref{3.4}) also admits the solution 
($0 < m < 1, c < 0$)
\be\label{A46}
u(\xi) = \frac{\sqrt{|c|}\beta}{\sqrt{2}}[\dn(\beta \xi,m) \pm 
\frac{\sqrt{m(1-m)}\sn(\beta \xi,m)}{\dn(\beta \xi, m)}]\,. 
\ee

It also follows that the focusing nmKdV-1 Eq. (\ref{3.6}) admits the solution
\be\label{A47}
u(\xi) = \frac{\beta}{2}[\dn(\beta \xi,m) \pm 
\frac{\sqrt{m(1-m)}\sn(\beta \xi,m)}{\dn(\beta \xi, m)}]\,. 
\ee

{\bf Solution XXVA}

One finds that 
\be\label{A48}
\phi(x) = \frac{[A+iB \sin(\beta x)]}{D+\cos(\beta x)}\,,
~~D > 1\,,
\ee
is an exact complex PT-invariant periodic solution with PT-eigenvalue 
$+1$ of $\phi^4$ Eq. (\ref{phieq}) provided
\be\label{A49}
2 b A^2 = -(D^2-1)\beta^2\,,~~2 b B^2 = -\frac{\beta^2}{2}\,,~~
a = \frac{\beta^2}{2}\,.
\ee

It then follows that the nKdV-1 Eq. (\ref{3.4}) also admits the solution 
($c < 0$)
\be\label{A50}
u(\xi) = \sqrt{|c|}{2}\beta \frac{\sqrt{D^2-1} \pm i\sin(\beta \xi)}
{D+\cos(\beta \xi)}\,,~~D > 1\,.
\ee

It also follows that the focusing nmKdV-1 Eq. (\ref{3.6}) admits the solution 
\be\label{A51}
u(\xi) = \frac{\beta}{2} \frac{\sqrt{D^2-1} \pm i\sin(\beta \xi)}
{D+\cos(\beta \xi)}\,,~~D > 1\,.
\ee

{\bf Solution XXVIA}

One finds that \be\label{A52}
\phi(x) = \frac{A\sqrt{m}\cn(\beta x,m)}{D+\dn(\beta x,m)}\,,~~D > 0\,,
\ee
is an exact periodic pulse solution of the symmetric $\phi^4$ Eq. (\ref{phieq}) 
provided
\be\label{A53}
0 < m < 1\,,~~D^2 = 1-m \,,~~2 b A^2 =  \beta^2\,,~~a 
= -(2-m)\frac{\beta^2}{2}\,.
\ee

Hence the corresponding superposed pulse solution of the nKdV-1 Eq. (\ref{3.4}) 
is ($c > 0$)
\be\label{A54}
u(\xi) = \frac{\sqrt{\frac{c}{2}} \beta \cn(\beta \xi,m)}{D+\dn(\beta \xi,m)}\,.
\ee

On the other hand, the corresponding superposed pulse solution of the 
defocusing nmKdV-1 Eq. (\ref{3.8}) is
\be\label{A55a}
u(\xi) = \frac{\frac{\beta}{2}\cn(\beta \xi,m)}{D+\dn(\beta \xi,m)}\,.
\ee

{\bf Solution XXVIIA}

We find  that 
\be\label{A55}
\phi(x) = \frac{A\dn(\beta x,m)}{D+\sn(\beta x,m)}\,,~~D > 1\,,
\ee
is an exact periodic pulse solution of the symmetric $\phi^4$ Eq. (\ref{phieq}) 
provided
\be\label{A56}
0 < m < 1\,,~~m D^2 = 1\,,~~2 bm A^2 =  -(1-m)\beta^2\,,~~a 
= (1+m)\frac{\beta^2}{2}\,.
\ee

Hence the corresponding pulse solution of the nKdV-1 Eq. (\ref{3.4}) 
is ($c < 0$)
\be\label{A57}
u(\xi) = \frac{\sqrt{\frac{(1-m)|c|}{2m}} \beta \dn(\beta \xi,m)}
{D+\sn(\beta \xi,m)}\,,~~0 < m < 1\,.
\ee

On the other hand, the corresponding pulse solution of the 
focusing nmKdV-1 Eq. (\ref{3.6}) is
\be\label{A58}
u(\xi) = \frac{\frac{\sqrt{1-m}\beta}{2\sqrt{m}}\dn(\beta \xi,m)}
{D+\sn(\beta \xi,m)}\,,~~0 < m < 1\,.
\ee

{\bf Solution XXVIIIA}

We find that 
\be\label{A59}
\phi(x) = \frac{A\dn(\beta x,m)+B\sqrt{m}\cn(\beta x,m)}{D+\sn(\beta x,m)}\,,
~~D > 1\,,
\ee
is an exact periodic superposed pulse solution of the symmetric $\phi^4$ 
Eq. (\ref{phieq}) provided
\be\label{A60}
0 < m < 1\,,~~2b A^2 = (D^2-1)\beta^2\,,~~2 b m B^2 =  -(m D^2-1)\beta^2\,,
~~a = (1+m)\frac{\beta^2}{2}\,.
\ee

Hence the corresponding pulse solution of the nKdV-1 Eq. (\ref{3.4}) 
is ($c < 0$)
\be\label{A61}
u(\xi) = \sqrt{\frac{|c|}{2}} \beta \frac{[\sqrt{D^2-1}\dn(\beta \xi,m)
+\sqrt{mD^2-1}\cn(\beta \xi,m)]}{D+\sn(\beta \xi,m)}\,,~~0 < m < 1\,.
\ee

On the other hand, the corresponding pulse solution of the 
focusing nmKdV-1 Eq. (\ref{3.6}) is
\be\label{A62}
u(\xi) = \frac{\beta}{2} \frac{[\sqrt{D^2-1}\dn(\beta \xi,m)
+\sqrt{mD^2-1}\cn(\beta \xi,m)]}{D+\sn(\beta \xi,m)}\,,~~0 < m < 1\,.
\ee

{\bf Solution XXIXA}

We find that 
\be\label{A63}
\phi(x) = \frac{[A\dn(\beta x,m)+iB\sqrt{m}\sn(\beta x,m)]}{D+\cn(\beta x,m)}\,,
~~D > 1\,,
\ee
is an exact complex PT-invariant periodic solution with PT-eigenvalue 
+$1$ of the $\phi^4$ Eq. (\ref{phieq}) provided
\be\label{A64}
2 b A^2 = -(D^2-1)\beta^2\,,~~2 m b B^2 = -(m D^2+1-m)\beta^2\,,~~
a = -(2m-1)\frac{\beta^2}{2}\,.
\ee

It then follows that the nKdV-1 Eq. (\ref{3.4}) also admits the superposed 
complex PT-invariant pulse solution ($c < 0$)
\be\label{A65}
u(\xi) = \sqrt{\frac{|c|}{2}} \beta \frac{[\sqrt{D^2-1}\dn(\beta \xi,m)
+i\sqrt{mD^2+1-m}\sn(\beta \xi,m)]}{D+\cn(\beta \xi,m)}\,,~~0 < m < 1\,.
\ee

It also follows that the focusing nmKdV-1 Eq. (\ref{3.6}) admits the superposed 
complex PT-invariant pulse solution 
\be\label{A66}
u(\xi) = \frac{\beta}{2} \frac{[\sqrt{D^2-1}\dn(\beta \xi,m)
+i\sqrt{mD^2+1-m}\sn(\beta \xi,m)]}{D+\cn(\beta \xi,m)}\,,~~0 < m < 1\,.
\ee

{\bf Solution XXXA}

We find that  that 
\be\label{A67}
\phi(x) = \frac{\sqrt{m}[A\cn(\beta x,m)+iB\sn(\beta x,m)]}{D+\dn(\beta x,m)}\,,
~~D > 0\,,
\ee
is an exact complex PT-invariant periodic solution with PT-eigenvalue 
+$1$ of the $\phi^4$ Eq. (\ref{phieq}) provided
\be\label{A68}
2 b A^2 = (1-D^2)\beta^2\,,~~2 b B^2 = (1-m- D^2)\beta^2\,,~~
a = -(2m-1)\frac{\beta^2}{2}\,.
\ee
Note that if $D^2 > 1$ then $b < 0$ while if $0 < D^2 < 1-m$, then $b > 0$.

It then follows that the nKdV-1 Eq. (\ref{3.4}) also admits the superposed 
complex PT-invariant pulse solution (with $D^2 > 1$ and hence $c < 0$)
\be\label{A65a}
u(\xi) = \sqrt{\frac{|c|}{2}} \beta \frac{[\sqrt{D^2-1}\cn(\beta \xi,m)
+i\sqrt{D^2-1+m}\sn(\beta \xi,m)]}{D+\cn(\beta \xi,m)}\,,~~0 < m < 1\,.
\ee
On the other hand if $0 < D^2 < 1-m$, then $c > 0$ and the solution is 
given by
\be\label{A66a}
u(\xi) = \sqrt{\frac{c}{2}} \beta \frac{[\sqrt{1-D^2}\cn(\beta \xi,m)
+i\sqrt{1-m-D^2}\sn(\beta \xi,m)]}{D+\cn(\beta \xi,m)}\,,~~0 < m < 1\,.
\ee

It also follows that the focusing nmKdV-1 Eq. (\ref{3.6}) admits the superposed 
complex PT-invariant pulse solution 
\be\label{A67a}
u(\xi) = \frac{\beta}{2} \frac{[\sqrt{D^2-1}\cn(\beta \xi,m)
+i\sqrt{D^2-1+m}\sn(\beta \xi,m)]}{D+\cn(\beta \xi,m)}\,,~~0 < m < 1\,.
\ee
On the other hand, the defocusing nmKdV-1 Eq. (\ref{3.8}) admits the 
complex PT-invariant pulse solution 
\be\label{A68a}
u(\xi) = \frac{\beta}{2} \frac{[\sqrt{1-D^2}\cn(\beta \xi,m)
+i\sqrt{1-m -D^2}\sn(\beta \xi,m)]}{D+\cn(\beta \xi,m)}\,,~~0 < m < 1\,.
\ee

In the limit $m = 1$, both the solutions XXIX and XXX go over to the complex 
PT-invariant hyperbolic solution with PT-eigenvalue $+1$, i.e.
\be\label{A69}
\phi(x) = \frac{[A\sech(\beta x)+iB\tanh(\beta x)]}{D+\sech(\beta x)}\,,
~~D > 0\,,
\ee
provided relations (\ref{A64}) with $m = 1$ are satisfied.

{\bf Solution XXXIA}

We also find that 
\be\label{A70}
\phi(x) = \frac{[A\sqrt{m}\sn(\beta x,m)+iB\dn(\beta x,m)]}{D+\cn(\beta x,m)}\,,
~~D > 1\,,
\ee
is an exact complex PT-invariant periodic solution with PT-eigenvalue 
$-1$ of the $\phi^4$ Eq. (\ref{phieq}) provided
\be\label{A71}
2 m b A^2 = (1-m+m D^2)\beta^2\,,~~2 b B^2 = (D^2-1)\beta^2\,,~~
a = -(2m-1)\frac{\beta^2}{2}\,.
\ee

It then follows that the nKdV-1 Eq. (\ref{3.4}) also admits the superposed 
complex PT-invariant pulse solution ($c > 0$)
\be\label{A72}
u(\xi) = \sqrt{\frac{|c|}{2}} \beta \frac{[\sqrt{m D^2+1-m}\sn(\beta \xi,m)
+i\sqrt{D^2-1}\dn(\beta \xi,m)]}{D+\cn(\beta \xi,m)}\,,~~0 < m < 1\,.
\ee

It also follows that the defocusing nmKdV-1 Eq. (\ref{3.8}) admits the 
superposed complex PT-invariant pulse solution 
\be\label{A73}
u(\xi) = \frac{\beta}{2} \frac{[\sqrt{mD^2+1-m}\sn(\beta \xi,m)
+i\sqrt{D^2-1}\dn(\beta \xi,m)]}{D+\cn(\beta \xi,m)}\,,~~0 < m < 1\,.
\ee

{\bf Solution XXXIIA}

We find that
\be\label{A74}
\phi(x) = \frac{\sqrt{m}[A\sn(\beta x,m)+iB\cn(\beta x,m)]}{D+\dn(\beta x,m)}\,,
~~D > 0\,,
\ee
is an exact complex PT-invariant periodic solution with PT-eigenvalue 
$-1$ of the $\phi^4$ Eq. (\ref{phieq}) provided
\be\label{A75}
2 b m A^2 = (D^2-1+m)\beta^2\,,~~2 b B^2 = (D^2-1)\beta^2\,,~~
a = -(2-m)\frac{\beta^2}{2}\,.
\ee
Note that if $D^2 > 1$ then $b > 0$ while if $0 < D^2 < 1-m$, then $b < 0$.

It then follows that the nKdV-1 Eq. (\ref{3.4}) also admits the superposed 
complex PT-invariant pulse solution (with $D^2 > 1$ and hence $c > 0$)
\be\label{A76}
u(\xi) = \sqrt{\frac{c}{2}} \beta \frac{[\sqrt{D^2-1+m}\sn(\beta \xi,m)
+i\sqrt{D^2-1}\cn(\beta \xi,m)]}{D+\dn(\beta \xi,m)}\,,~~0 < m < 1\,.
\ee
On the other hand if $0 < D^2 < 1-m$, then $c < 0$ and the solution is 
given by
\be\label{A77}
u(\xi) = \sqrt{\frac{|c|}{2}} \beta \frac{[\sqrt{1-m - D^2}\sn(\beta \xi,m)
+i\sqrt{1-D^2}\cn(\beta \xi,m)]}{D+\dn(\beta \xi,m)}\,,~~0 < m < 1\,.
\ee

It also follows that the defocusing nmKdV-1 Eq. (\ref{3.8}) admits the 
superposed complex PT-invariant pulse solution 
\be\label{A78}
u(\xi) = \frac{\beta}{2} \frac{[\sqrt{D^2-1+m}\sn(\beta \xi,m)
+i\sqrt{D^2-1}\cn(\beta \xi,m)]}{D+\dn(\beta \xi,m)}\,,~~0 < m < 1\,.
\ee
On the other hand, the focusing nmKdV-1 Eq. (\ref{3.6}) admits the 
complex PT-invariant pulse solution 
\be\label{A79}
u(\xi) = \frac{\beta}{2} \frac{[\sqrt{1-m-D^2}\sn(\beta \xi,m)
+i\sqrt{1-D^2}\cn(\beta \xi,m)]}{D+\dn(\beta \xi,m)}\,,~~0 < m < 1\,.
\ee

In the limit $m = 1$, both the solutions XXXI and XXXII go over to the complex 
PT-invariant hyperbolic solution with PT-eigenvalue $-1$, i.e.
\be\label{A80}
\phi(x) = \frac{[A\tanh(\beta x)+iB\sech(\beta x)]}{D+\sech(\beta x)}\,,
~~D > 0\,,
\ee
provided relations (\ref{A71}) with $m = 1$ are satisfied.

\end{document}